\documentclass[a4paper,fleqn,usenatbib]{mnras}

\usepackage{txfonts}
\usepackage[T1]{fontenc}
\usepackage{ae,aecompl}
\usepackage{xcolor}

\usepackage{hyperref}

\usepackage{graphicx}	% Including figure files
\usepackage{pdflscape}	% Extra maths symbols
\usepackage[english]{babel}
\usepackage{placeins}
\newcommand{\ethylcyanide}{CH$_3$CH$_2$CN}

\newcommand{\methylcyanide}{CH$_3$CN}

\newcommand{\hcccn}{HC$_3$N}  
\newcommand{\methanol}{CH$_3$OH}

\newcommand{\lsun}{$L_\odot$}
\newcommand{\msun}{$M_\odot$}

\title[G11.92--0.61 MM1: A Keplerian disc around a massive young proto-O star]{G11.92--0.61 MM1: A Keplerian disc around a massive young proto-O star}

\author[J. D. Ilee et al.]
{J.~D.~Ilee$^1$\thanks{Contact e-mail: \href{mailto:jdilee@ast.cam.ac.uk}{jdilee@ast.cam.ac.uk}},
C.~J.~Cyganowski$^2$,
P.~Nazari$^2$, 
T.~R.~Hunter$^3$,
C.~L.~Brogan$^3$,
\newauthor 	% starts a new line in the author environment
D.~H.~Forgan$^2$ and  
Q.~Zhang$^4$\vspace{0.2cm} \\
\\
$^{1}$Institute of Astronomy, Madingley Road, Cambridge CB3 0HA, UK\\
$^{2}$SUPA, School of Physics \& Astronomy, University of St Andrews, North Haugh, St Andrews, Scotland, KY16 9SS, UK\\
$^{3}$NRAO, 520 Edgemont Rd, Charlottesville, VA 22903, USA,\\
$^{4}$Harvard-Smithsonian Center for Astrophysics, Cambridge, MA 02138, USA
}

\date{Accepted 2016 July 29. Received 2016 July 29; in original form 2016 April 22}

\pubyear{2016}

\begin{document}
\label{firstpage}
\pagerange{\pageref{firstpage}--\pageref{lastpage}}
\maketitle

\begin{abstract}
    The formation process of massive stars is not well understood, and
    advancement  in our  understanding benefits  from high  resolution
    observations  and  modelling  of  the  gas  and  dust  surrounding
    individual high-mass  (proto)stars.  Here we  report sub-arcsecond
    ($\lesssim$1550\,au) resolution observations  of the young massive
    star  G11.92--0.61 MM1  with  the SMA  and VLA.   Our  1.3 mm  SMA
    observations  reveal  consistent  velocity  gradients  in  compact
    molecular  line   emission  from   species  such   as  CH$_{3}$CN,
    CH$_{3}$OH,  OCS,  HNCO,  H$_{2}$CO, DCN  and  CH$_{3}$CH$_{2}$CN,
    oriented   perpendicular   to  the   previously-reported   bipolar
    molecular  outflow  from  MM1.    Modelling  of  the  compact  gas
    kinematics  suggests a  structure undergoing  rotation around  the
    peak of the  dust continuum emission.  The  rotational profile can
    be well  fit by  a model  of a  Keplerian disc,  including infall,
    surrounding  an  enclosed  mass of  $\sim$30--60\,M$_{\odot}$,  of
    which 2--3\,M$_{\odot}$ is attributed to the disc.  From modelling
    the  CH$_{3}$CN  emission,  we   determine  that  two  temperature
    components,  of   $\sim$  150\,K  and  230\,K,   are  required  to
    adequately  reproduce  the  spectra.   Our  0.9  and  3.0\,cm  VLA
    continuum data  exhibit an  excess above  the level  expected from
    dust  emission;   the  full   centimetre-submillimetre  wavelength
    spectral energy distribution of MM1  is well reproduced by a model
    including dust  emission, an  unresolved hypercompact  H\,{\sc ii}
    region, and  a compact ionised  jet.  In combination,  our results
    suggest that MM1  is an example of a massive  proto-O star forming
    via disc accretion, in a similar way to that of lower mass stars.
\end{abstract}


%%%%%%%%%%%%%%%%%%%%%%%%%%%%%%%%%%%%%%%%%%%%%%%%%%

%%%%%%%%%%%%%%%%% BODY OF PAPER %%%%%%%%%%%%%%%%%%

\section{Introduction}
\label{sec:introduction}

Massive  stars   ($M_{\star}  >  8\,M_{\odot}$)  have   a  very  short
pre-main-sequence  phase  \citep[$\sim$10$^{4}$   --  a  few  $\times$
  10$^{5}$\,years,  e.g.][]{davies_2011,mottram_2011},   meaning  that
they spend the  entirety of their formation stages  deeply embedded in
their parent  molecular clouds.  Such short  formation timescales also
mean that  there are far fewer  examples of young, massive  stars when
compared with  their lower mass counterparts.   This inherent scarcity
means that young  massive stars are, on average, located  in much more
distant  star  forming  regions  ($>1$\,kpc).  All  of  these  factors
contribute to the extreme difficulty  of observing young massive stars
directly.  As   a  result,  their  formation   mechanisms  are  poorly
understood.

\smallskip

In particular, the process by  which young massive stars accrete their
high masses  is not known.   There is  little time to  accumulate such
large masses  during their  short formation timescales,  and accretion
onto  the central  object may  be halted  or substantially  reduced by
energetic feedback processes (such as high radiation pressures, strong
stellar  winds and  ionising radiation).   Models have  suggested that
channelling  material  through  a  circumstellar  accretion  disc  can
overcome   these    feedback   mechanisms   \citep[e.g.][]{yorke_2002,
  krumholz_2009,      kuiper_2010,     kuiper_2011,      harries_2014,
  klassen_2016}.

\smallskip

Observationally, however,  it is  not yet clear  whether circumstellar
accretion discs surround massive young  stellar objects (MYSOs) of all
masses and evolutionary stages.   In particular, convincing candidates
for Keplerian discs around embedded O-type (proto)stars are proving to
be      particularly      elusive     \citep[e.g.][and      references
  therein]{cesaroni_2007,wang_2012,beltran_2016}.     Infrared    (IR)
interferometry  \citep{kraus_2010,   boley_2013}  and  high-resolution
near-infrared      spectroscopy     \citep{bik_2004,      davies_2010,
  wheelwright_2010,  ilee_2013,  ilee_2014}  have  revealed  discs  on
scales of less than $1000$\,au  around MYSOs, but these techniques are
limited to  relatively evolved, IR-bright  objects.  Longer-wavelength
interferometric observations allow  access to the circum(proto)stellar
environments of  less evolved,  more embedded  MYSOs, but  often probe
larger spatial scales.  In many  cases, velocity gradients detected in
millimetre and centimetre-wavelength molecular line observations trace
`toroids'  ---  large-scale  (1000s  to  $\gtrsim$10,000\,au),  massive
($\mathrm{M}_{\mathrm{toroid}}        \ge        \mathrm{M}_{\star}$),
non-equilibrium   rotating   structures  \citep[e.g.][and   references
  therein]{cesaroni_2005_apss,cesaroni_2006,cesaroni_2007,beuther_2008,beltran_2011,cesaroni_2011,johnston_2014}. The
clustered nature of massive star formation also complicates the search
for accretion disc  candidates, as multiplicity and  rotation can both
produce                       velocity                       gradients
\citep[c.f.][]{patel_2005,brogan_2007,comito_2007},   and  very   high
angular resolution  observations are  required to  distinguish between
these scenarios.

\smallskip

Searching for  small-scale Keplerian  accretion discs  around embedded
distant  MYSOs  requires  sub-arcsecond-resolution  observations  with
(sub)millimetre  interferometers.  Recent  studies of  this type  have
yielded  a   handful  of  candidate  Keplerian   discs  around  O-type
(proto)stars
\citep{jimenez-serra_2012,qiu_2012,wang_2012,hunter_2014,johnston_2015,zapata_2015},
as well as adding to the sample of good candidates for Keplerian discs
around                       B-type                       (proto)stars
\citep[e.g.][]{sanchez_2013,beltran_2014,cesaroni_2014}.

\smallskip

In this paper, we  report sub-arcsecond resolution Submillimeter Array
(SMA) observations of a candidate disc around a high-mass (proto)star,
G11.92--0.61  MM1 (hereafter  MM1), identified  in the  course of  our
studies       of       GLIMPSE      Extended       Green       Objects
\citep[EGOS;][]{cyganowski_2008}.   MM1   is  one  of   three  compact
millimetre continuum  cores detected in  thermal dust emission  in our
initial   SMA   1.3\,mm   observations   of   the   EGO   G11.92--0.61
\citep[][resolution  $\sim$2.4\arcsec]{cyganowski_2011sma},  which  is
located in  an infrared dark  cloud (IRDC) $\sim$1\arcmin\/ SE  of the
more evolved massive star-forming region \emph{IRAS} 18110--1854.  The
three members of the G11.92--0.61  (proto)cluster are only resolved at
(sub)millimetre            and           longer            wavelengths
\citep{cyganowski_2011sma,cyganowski_2014};  the  total luminosity  of
the             region              is             $\sim$10$^{4}$\lsun
\citep{cyganowski_2011sma,moscadelli_2016}.    MM1  exhibits   copious
molecular      line     emission      from     hot-core      molecules
\citep{cyganowski_2011sma, cyganowski_2014}  and is coincident  with a
6.7 GHz  Class II CH$_{3}$OH maser  \citep{cyganowski_2009} and strong
H$_{2}$O                                                        masers
\citep{hofner_1996,breen_2011,sato_2014,moscadelli_2016},          all
indicative of the  presence of a massive (proto)star.  MM1  also has a
weak ($\lesssim$1\,mJy) cm-wavelength counterpart, CM1, first detected
at 1.3\,cm by  \citet{cyganowski_2011vla,cyganowski_2014} and recently
also at 4.8  and 2.3\,cm by \citet{moscadelli_2016} as part  of a Very
Large Array (VLA) survey of BeSSeL H$_{2}$O maser sources.

\smallskip

\citet{cyganowski_2011sma}   detected   a  single   dominant   bipolar
molecular outflow towards  the G11.92--0.61 millimetre (proto)cluster,
driven   by  MM1   and   traced   by  well-collimated,   high-velocity
$^{12}$CO(2--1) and  HCO$^{+}$(1--0) emission.   Near-infrared H$_{2}$
emission  \citep{lee_2012,lee_2013}  and  44\,GHz Class  I  CH$_{3}$OH
masers  \citep{cyganowski_2009} also  trace shocked  outflow gas.   On
small  ($<  1000$\,au)  scales,  the  H$_{2}$O  maser  proper  motions
indicate a collimated  NE-SW flow, consistent with  the orientation of
the    large-scale   molecular    outflow   \citep{cyganowski_2011sma,
  cyganowski_2014,          moscadelli_2016}.          In          our
$\sim$2.4\arcsec-resolution SMA data,  the compact, hot-core molecular
line  emission  associated  with  MM1  displays  a  velocity  gradient
oriented   roughly  perpendicular   to  the   outflow  axis,   leading
\citet{cyganowski_2011sma} to suggest an unresolved disc as a possible
explanation,  but   requiring  higher  angular  resolution   data  for
confirmation.

\smallskip

In this paper, we  present sub-arcsecond-resolution line and continuum
observations  of   G11.92-0.61  obtained  with  the   highest  angular
resolution possible  with the SMA  at 1.3  mm.  Together with  new VLA
subarcsecond-resolution centimetre continuum  observations, we use the
SMA data to study the molecular gas kinematics of MM1 and to constrain
the nature of the central  source.  Our observations are summarised in
Section~\ref{sec:obs}, and  our results  in Section~\ref{sec:results}.
Section~\ref{sec:discussion} presents our  modelling of the kinematics
and   physical  properties   of  the   candidate  disc   and  of   the
centimetre-(sub)millimetre  wavelength   emission  from   the  central
source, and discusses our results.   Our conclusions are summarised in
Section~\ref{sec:conclusions}.  We  adopt the maser  parallax distance
to    G11.92--0.61     of    3.37$^{+0.39}_{-0.32}$\,kpc    throughout
\citep{sato_2014}

\begin{table*}
	\begin{minipage}{0.9\textwidth}
	\centering
	\caption{Parameters of the SMA and VLA observations.}
	\label{tab:obs}
	\begin{tabular}{lcccc} 
	\hline
    Parameter & SMA 1.3\,mm & SMA 0.88\,mm & VLA 3\,cm & VLA 0.9\,cm \\
	\hline
	Observing date (UT) & 2011 Aug 28 &  2011 Aug 19 & 2015 Jun 25     & 2015 Feb 9-10 \\
	Project code        & 2011A-S076  &  2011A-S076  & 15A-232 & 15A-232\\ % removed TRH 2016-06-02, 10A-155 \\
	Configuration       & Very Extended & Extended &    A &         B \\
	Phase Center (J2000): & & & \\
	~~~~~R.A. & 18$^{\rm h}$13$^{\rm m}$58$\fs$10 & 18$^{\rm h}$13$^{\rm m}$58$\fs$10 & 18$^{\rm h}$13$^{\rm m}$58$\fs$10 &  18$^{\rm h}$13$^{\rm m}$58$\fs$10 \\
	~~~~~Dec.                     & $-$18$\degr$54\arcmin16\farcs7 &  $-$18$\degr$54\arcmin16\farcs7 & $-$18$\degr$54\arcmin16\farcs7 & $-$18$\degr$54\arcmin16\farcs7 \\
	Primary beam size (FWHP) & 52\arcsec &  34\arcsec &  4\arcmin\/      & 1.3\arcmin\/ \\
	Frequency coverage: & & & & \\
	Lower band (LSB) center & 218.9 GHz & 335.6 GHz &  9 GHz            &  31 GHz\\  
	Upper band (USB) center & 230.9 GHz & 347.6 GHz &  11 GHz           &  35 GHz \\
	Bandwidth         & 2 $\times$ 4 GHz  & 2 $\times$ 4 GHz & 2 $\times$ 2.048 GHz & 4 $\times$ 2.048 GHz \\
	Subbands          & n/a            & n/a                 &  2 $\times$ 16        & 4 $\times$ 16       \\
	Channel spacing   & 0.8125 MHz &   0.8125 MHz  &  1 kHz & 1 kHz \\
	                  &            &               & (30\,km\,s$^{-1}$) & (9\,km\,s$^{-1}$) \\
	Gain calibrator(s) & J1733--130, J1924--292 & J1733--130, J1924--292& J1832--2039 & J1832--2039 \\
	Bandpass calibrator & 3C84 & 3C84 & J1924--2914 & J1924--2914 \\
	Flux calibrator & Callisto$^{a}$ & Callisto$^{a}$ & J1331+3030 & J1331+3030 \\
	Synthesised beam$^{b}$         &  0\farcs57$\times$0\farcs37 & 0\farcs80$\times$0\farcs70  & 0\farcs30$\times$0\farcs17  & 0\farcs31$\times$0\farcs17  \\
	                         &  ($\mathrm{P.A.}=30^{\circ}$) & ($\mathrm{P.A}.=54^{\circ}$) & ($\mathrm{P.A.}=0^{\circ}$) & ($\mathrm{P.A.}=-5^{\circ}$) \\
	Continuum rms noise$^{b}$ &  0.7 mJy beam$^{-1}$ & 3 mJy beam$^{-1}$ & 6.1 $\mu$Jy beam$^{-1}$  & 7.6 $\mu$Jy beam$^{-1}$ \\
	Spectral line rms noise$^{c}$ & 23 mJy beam$^{-1}$ & 55 mJy beam$^{-1}$ ($^{12}$CO) & n/a & n/a \\
	\hline
	\end{tabular}
        \begin{flushleft}
	\small{$a$:        Using       Butler-JPL-Horizons        2012
          models.}\\  \small{$b$:  SMA: for
          combined LSB+USB continuum image (Briggs weighting, robust =
          0.5).}\\    \small{$c$:    Typical    rms    per    channel;
          Hanning-smoothed.  For $^{12}$CO(3--2),  rms is per smoothed
          3\,km\,s$^{-1}$              channel              \citep[see
            also][]{cyganowski_2014}.}\\
        \end{flushleft}
	\end{minipage}
\end{table*}

\section{Observations}
\label{sec:obs}

\subsection{Submillimeter Array (SMA) \label{sec:obs_sma}}

The SMA 1.3\,mm Very Extended configuration (VEX) dataset is described
in   \citet{cyganowski_2014},  which   uses   these   data  to   study
G11.92--0.61  MM2,   a  candidate  massive  prestellar   core  located
$\sim$7\farcs2 from MM1 in the G11.92--0.61 millimetre (proto)cluster.
The  SMA  1.3\,mm  observations  were  taken  in  good  weather,  with
$\tau_{\rm 225 GHz}\sim$0.05 and system temperatures at source transit
T$_{\rm  sys}\sim$90\,K;   eight  antennas  were  available   for  the
observations.    Key   observational    parameters   are   listed   in
Table~\ref{tab:obs}.  The 1.3\,mm line data were resampled to a common
velocity resolution of 1.12\,km\,s$^{-1}$, then Hanning-smoothed.  The
projected    baselines   ranged    from   $\sim$20--394    k$\lambda$,
corresponding to  a largest angular  scale (for sensitivity  to smooth
emission) of $\sim$9\arcsec\/ \citep[see also][]{cyganowski_2014}.  We
carefully identified the  line-free portions of the  spectrum in order
to perform  continuum subtraction  and to  generate a  continuum image
with  minimal  line contamination.   The  effective  bandwidth of  the
continuum  image is  approximately two  thirds of  the total  observed
bandwidth.  All measurements  were made from images  corrected for the
primary beam response.

\smallskip

To measure the position angle  of the bipolar molecular outflow driven
by  MM1, we  make  use of  the $^{12}$CO(3--2)  line  included in  the
0.88\,mm SMA dataset presented in \citet{cyganowski_2014}.  Details of
the SMA 0.88\,mm observations  are included in Table~\ref{tab:obs} for
completeness; here, we consider only the $^{12}$CO(3--2) data.

\subsection{Jansky Very Large Array (VLA)}

In 2015,  we observed G11.92--0.61 with  the NRAO Karl G.  Jansky Very
Large Array  (VLA) in two  bands (X and  Ka, 3\,cm and  0.9\,cm) under
project code 15A--232.  The observations were taken with dual circular
polarization,  with on-source  observing times  of 39  minutes and  86
minutes  at  3 and  0.9\,cm,  respectively.   Further details  of  the
observations  are  given in  Table~\ref{tab:obs}.   Due  to the  broad
bandwidth of  the datasets,  we imaged  them using  2 Taylor  terms to
account  for  the  spectral  index   of  the  emission  and  performed
phase-only  self-calibration. Due  to  the proximity  of the  cometary
ultra-compact     (UC)     H\,{\sc     ii}     region     G11.94--0.62
\citep{Wood1989apjs} about  $1'$ to the north-northeast,  we needed to
include that  source in the model.   The X-band image was  made with a
minimum uv distance of 1300~k$\lambda$  in order to remove ripple from
G11.94--0.62    in     the    vicinity    of     G11.92--0.61.     Our
previously-published       VLA       observations      in       K-band
\citep[1.2\,cm;][]{cyganowski_2014}  are also  used  in modelling  the
spectral energy distribution in Section~\ref{sec:ff_sed}.

\begin{figure*}
    \includegraphics[width=\textwidth]{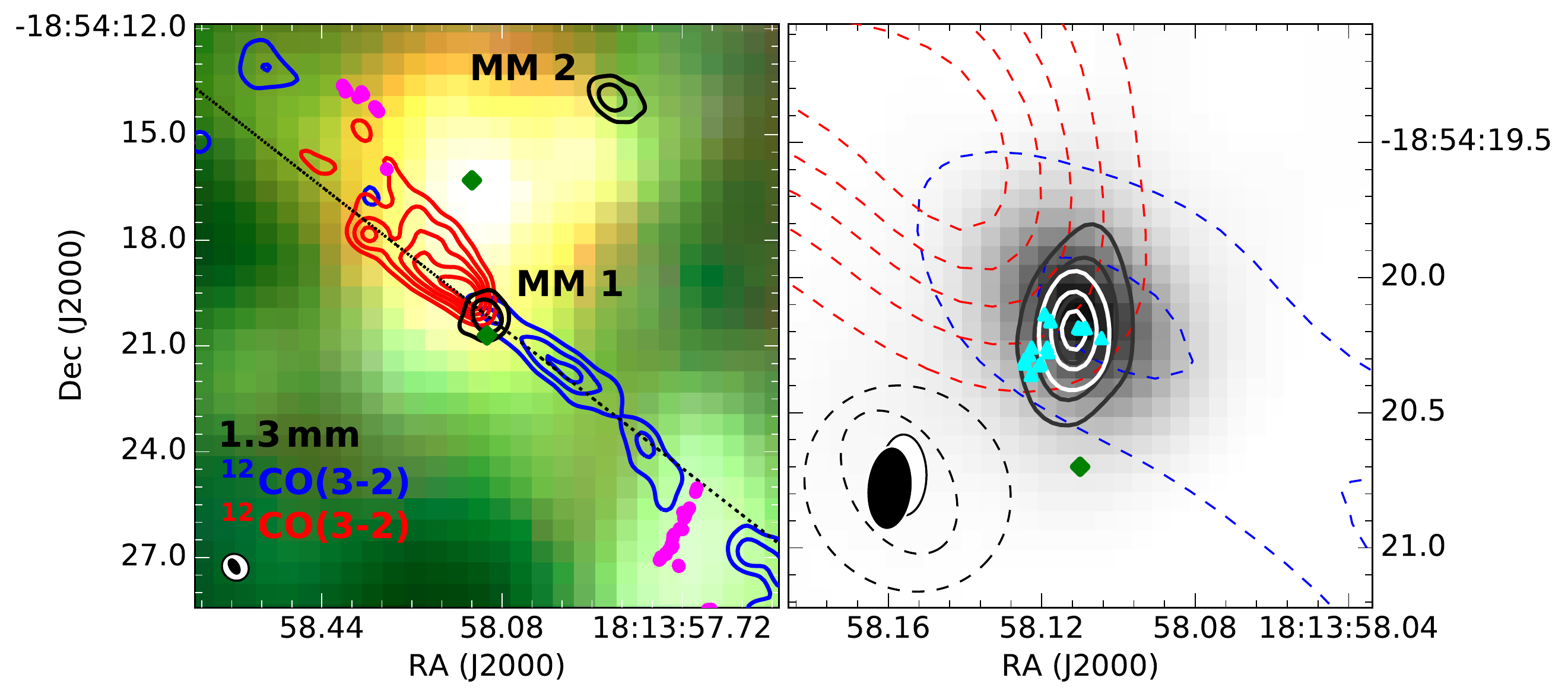}
    \caption{\textit{Left:}  Contours  of  SMA 1.3\,mm  VEX  continuum
      (black) and blue/redshifted $^{12}$CO(3--2) emission overlaid on
      a three-color \emph{Spitzer} image (RGB: 8.0, 4.5, 3.6\,$\mu$m).
      The  outflow  position angle  ($52\degr$)  is  indicated by  the
      dotted          black         line          (see         Section
      \ref{sec:results_compactlines}). The field of view is centred on
      MM1,  and  MM2 is  visible  $\sim$7\arcsec\/  to the  northwest.
      Positions of 44\,GHz Class I CH$_{3}$OH masers (magenta circles)
      and 6.7\,GHz  Class II  CH$_{3}$OH masers (green  diamonds) from
      \citet{cyganowski_2009}   are   marked.    The  6.7\,GHz   maser
      $\sim$5\arcsec\/ north of MM1  is coincident with the millimetre
      continuum   source   MM3    \citep[][undetected   in   the   VEX
      observations]{cyganowski_2011sma}.   \textit{Right:} Zoomed view
      of MM1, showing the SMA 1.3\,mm VEX continuum (greyscale, linear
      from   0  to   0.1\,Jy\,beam$^{-1}$),   $^{12}$CO(3-2)  emission
      (blue/red  dashed contours)  and  6.7 GHz  maser (diamond).   In
      addition,  VLA 3.0\,cm  (white) and  0.9\,cm (black)  contours are
      overlaid, along with positions  of 22\,GHz H$_{2}$O masers (cyan
      triangles, \citealt{moscadelli_2016}).  Beams are shown at lower
      left  in  each panel.   Levels:  1.3\,mm: (5,25)$\sigma$,  where
      $\sigma=0.7$\,mJy   \,beam$^{-1}$;  3.0\,cm:  (5,15,25)$\sigma$,
      where        $\sigma=6.1$\,$\mu$Jy\,beam$^{-1}$;        0.9\,cm:
      (5,15,50)$\sigma$,   where   $\sigma=7.6$\,$\mu$Jy\,beam$^{-1}$;
      $^{12}$CO: 0.8\,Jy\,beam$^{-1}$\,km\,s$^{-1}$ $\times$ (5,10,15) (blue),
      $\times$ (5,10,15,20,25) (red).}
    \label{fig:outflow}
\end{figure*}

\section{Results}                                                    
\label{sec:results}                                                                       

\subsection{Continuum emission}
\label{sec:cont_results}

Figure~\ref{fig:outflow}  shows  our   new  VLA  centimetre-wavelength
continuum  images  overlaid   on  the  1.3\,mm  SMA   VEX  image  from
\citet{cyganowski_2014}; observed centimetre  continuum properties for
CM1  are listed  in Table~\ref{tab:cm_prop}.   The continuum  emission
from CM1 is compact at all wavelengths.  The fitted centroid positions
for  CM1  in  the  3.0   and  0.9\,cm  images  are  consistent  within
$<$0\farcs01, and  are within $\sim$0\farcs05  of the 1.3 mm  MM1 peak
\citep[fitted centroid position,  Table 2 of][]{cyganowski_2014}.  CM1
is unresolved  in our deep VLA  images, indicating that the  source of
centimetre-wavelength     emission     is     smaller     than     our
$\sim$1000$\times$570\,au beam.  This result is consistent with recent
VLA  observations  at  4.8,  2.3 and  1.4\,cm  (resolution  0\farcs32,
0\farcs19,  and  0\farcs1, respectively):  in  all  cases, the  fitted
source  size   is  less  than   the  size  of  the   synthesised  beam
\citep[][their Table 3]{moscadelli_2016}.

\begin{table*}
	\begin{minipage}{0.955\textwidth}
	\centering
	\caption{Observed continuum emission properties.}
	\label{tab:cm_prop}
	\begin{tabular}{cccccccc} 
          \hline
          			& \multicolumn{2}{c}{J2000} 			& Peak 		 	& Integrated 			& 	 			& 		&   	\\
	 Obs. $\lambda$		& \multicolumn{2}{c}{Coordinates$^{a}$}		& Intensity$^{a}$	& Flux Density$^{a}$ 		& Size$^{a}$			& Size$^{a}$	& P.A.	\\

           			& $\alpha (\mathrm{h~m~s})$ & $\delta (\degr~'~'')$  	& (mJy beam$^{-1}$) 	& (mJy) 			& (\arcsec $\times$ \arcsec)	& (au $\times$ au) 	&  ($^{\circ}$) 	\\ 
          \hline
          1.3\,mm			& 18 13 58.1099 	& $-$18 54 20.141		& $90.0\pm1.0$		& $138.0\pm2.0$			& $0.34\pm0.01 \times 0.29\pm0.01$ & $1150 \times 960$ 	& $125\pm7$	\\
          0.9\,cm 			& 18 13 58.1108 	& $-$18 54 20.185 		& $0.548\pm0.011$ 		& $0.715\pm0.023$ 			& $0.16\pm0.02 \times 0.10\pm0.01$ & $550 \times 330$ 	& $160\pm12$ 	\\
          3.0\,cm 			& 18 13 58.1113 	& $-$18 54 20.191 		& $0.203\pm0.009$ 		& $0.161\pm0.014$ 			& \ldots & \ldots & \ldots   						\\ 

	\hline
	\end{tabular}
	\begin{flushleft}
          \small{$a$: From two-dimensional Gaussian fitting; ``size'' is
            deconvolved source size.  If  no size is given, the source
            could  not be  deconvolved  from the  beam,  and thus  its
            appearance cannot  be distinguished  from that of  a point
            source.  Statistical uncertainties are indicated
            by   the  number   of  significant   figures.}\\
	\end{flushleft}
\end{minipage}
\end{table*}

\subsection{Compact molecular line emission: MM1}
\label{sec:results_compactlines}

In  our SMA  1.3\,mm  VEX observations,  the  molecular line  emission
detected towards  G11.92-0.61 consists primarily of  compact emission,
coincident with the MM1 (sub)millimetre continuum source, from species
characteristic   of   hot   cores  \citep[including   CH$_3$CN,   OCS,
  CH$_3$CH$_2$CN,     and    HC$_3$N;     see     also    Figure     2
  of][]{cyganowski_2014}.  From the 8\,GHz  of bandwidth observed with
the SMA at 1.3\,mm, we identified  31 lines from 10 different chemical
species that are potentially strong and unblended enough to be used to
study the  kinematics of MM1.   Details for  these lines are  given in
Table~\ref{tab:linelist}, ordered by  decreasing E$_{\rm upper}$.  For
all  lines  included  in   Table~\ref{tab:linelist},  we  created  and
inspected  integrated intensity  (moment  0) and  velocity (moment  1)
maps.  The moment 1 maps were constructed using an intensity threshold
of  5$\sigma$,   and  the   velocity  extent   was  chosen   to  avoid
contamination from nearby lines where possible.

\smallskip

\begin{table*}
  \begin{minipage}{0.715\textwidth}
    \begin{center}
\caption{Properties of spectral lines used in kinematic analysis.}
\label{tab:linelist}
\begin{tabular}{lccccc}
\hline
Species				& Transition			&  Frequency		& E$_{\mathrm{upper}}$ 	 & Catalog$^{a,b}$	& Notes$^{c}$		\\
				&	    			&  (GHz)			& (K)		      	&  	        &		     					\\	
\hline
\methylcyanide\          	& $J=12$--11, $K=8$           	& 220.47581		& 525.6  		& JPL 		&						\\
\hcccn\ v$_{\rm 7}=$1      	& $J=24$--23 l=1f          	& 219.17376  		& 452.3  		& CDMS 		&							\\
\hcccn\ v$_{\rm 7}=$1      	& $J=24$--23 l=1e           	& 218.86080  		& 452.1  		& CDMS 		&							\\
\methanol\/ ($A$)        	& 18$_{3,16}$--17$_{4,13}$   	& 232.78350 		& 446.5 		& CDMS 		&							\\
\methylcyanide\          	& $J=12$--11, $K=7$           	& 220.53932 		& 418.6 		& JPL 		& PA, DT\_ALL						\\
\methanol\/ ($E$)        	& 15$_{4,11}$--16$_{3,13}$   	& 229.58907  		& 374.4 		& CDMS 		&							\\
\methanol\/ v$_t$=1 ($A$)	& 6$_{1,5}$--7$_{2,6}$       	& 217.29920 		& 373.9 		& CDMS 		& PA, DT\_CH$_3$OH			\\
\methylcyanide\         	& $J=12$--11, $K=5$            	& 220.64108 		& 247.4 		& JPL 		&							\\
HNCO$^{d}$                    	& 10$_{2,9}$--9$_{2,8}$       	& 219.73385 		& 228.3 		& CDMS		&							\\
\methylcyanide\         	& $J=12$--11, $K=4$            	& 220.67929 		& 183.1 		& JPL 		&		PA, DT\_ALL					\\
\methanol\/ ($A$)       	& 10$_{2,8}$--9$_{3,7}$       	& 232.41859 		& 165.4 		& CDMS 		& PA, DT\_CH$_3$OH			\\
\methanol\/ ($A$)       	& 10$_{2,9}$--9$_{4,6}$     	& 231.28110 		& 165.3 		& CDMS 		&	PA, DT\_CH$_3$OH		\\
\ethylcyanide\          	& 27$_{0,27}$--26$_{0,26}$  	& 231.99041 		& 157.7 		& CDMS 		&							\\
\ethylcyanide\          	& 26$_{2,25}$--25$_{2,24}$  	& 229.26516 		& 154.0 		& CDMS 		&							\\
\ethylcyanide\          	& 26$_{1,25}$--25$_{1,24}$  	& 231.31042 		& 153.4 		& CDMS 		&							\\
\ethylcyanide\                  & 25$_{2,24}$--24$_{2,23}$ 	& 220.66092 		& 143.0 		& CDMS 		&	PA, DT\_ALL					\\	
\methylcyanide\         	& $J=12$--11, $K=3$          	& 220.70902 		& 133.2 		& JPL 		&		PA, DT\_ALL					\\
\hcccn\                 	& $J=24$--23               	& 218.32472  		& 131.0 		& CDMS 		&							\\
CH$_3$OCHO ($A$)$^{e}$ 		& 20$_{1,20}$--19$_{1,19}$ 	& 216.96590 		& 111.5 		& JPL 		&							\\	
OCS                     	& $J=19$--18               	& 231.06099  		& 110.9 		& CDMS 		&	PA, DT\_ALL					\\
HNCO                    	& 10$_{1,10}$--9$_{1,9}$    	& 218.98101  		& 101.1 		& CDMS 		&	PA, DT\_ALL					\\
\methylcyanide\         	& $J=12$--11, $K=2$          	& 220.73026  		& 97.4  		& JPL 		&	PA, DT\_ALL						\\
\methanol\/ ($E$)       	& 8$_{0,8}$--7$_{1,6}$    	& 220.07849  		& 96.6 			& CDMS 		&	PA, DT\_CH$_3$OH			\\
\methanol\/ ($E$)$^{f}$       	& 8$_{-1,8}$--7$_{0,7}$   	& 229.75876  		& 89.1 			& CDMS 		&		PA, DT\_CH$_3$OH		\\
H$_{2}$CO               	&  3$_{2,1}$--2$_{2,0}$    	& 218.76007  		& 68.1 			& CDMS 		&		PA, DT\_ALL					\\
HNCO                   		& 10$_{0,10}$--9$_{0,9}$    	& 219.79827  		& 58.0 			& CDMS 		&							\\
\methanol\/ ($E$)      		& 5$_{1,4}$--4$_{2,2}$    	& 216.94560 		& 55.9 			& CDMS 		&							\\
\methanol\/ ($E$)      		& 3$_{-2,2}$--4$_{-1,4}$   	& 230.02706 		& 39.8 			& CDMS 		&	PA, DT\_CH$_3$OH		\\
SO                     		&  6$_{5}$--5$_{4}$       	& 219.94944 		& 35.0 			& CDMS 		&							\\
H$_{2}$CO              		&  3$_{0,3}$--2$_{0,2}$    	& 218.22219 		& 21.0 			& CDMS 		&							\\
DCN                   		& $J=3$--2                	& 217.23854 		& 20.9 			& CDMS 		&	PA, DT\_ALL						\\
\hline
\end{tabular}
\end{center}   
\small{$^{a}$ CDMS = \url{http://www.astro.uni-koeln.de/cgi-bin/cdmssearch}}\\
\small{$^{b}$ JPL = \url{http://spec.jpl.nasa.gov/ftp/pub/catalog/catform.html}}\\
\small{$^{c}$ PA: line included in calculation of median position angle (Section~\ref{sec:results_compactlines}); DT\_ALL: Potentially `disc tracing' --- results of centroid plot included in the kinematic fitting, all lines excluding CH$_{3}$OH (Section~\ref{sec:kinematics}); DT\_CH$_3$OH: Potentially `disc tracing' --- results of centroid plot included in the kinematic fitting, CH$_{3}$OH lines (Section~\ref{sec:kinematics})} \\
\small{$^{d}$ Blended with the HNCO 10$_{2,8}-$9$_{2,7}$ line at 219.73719\,GHz, which has the same E$_{\mathrm{upper}}$ and CDMS intensity; the
velocity separation from this reference transition is -4.56 km s$^{-1}$.}\\
\small{$^{e}$ Blended with three other CH$_{3}$OCHO transitions with the same line intensity and E$_{\mathrm{upper}}$; the velocity separations
from this reference transition are +1.57, -0.48, and -2.10\,km\,s$^{-1}$.}\\  
\small{$^{f}$ This line exhibits thermal emission towards MM1, though maser emission is observed towards the G11.92$-$0.61 outflow lobes
\citep{cyganowski_2011sma}.}\\ 
\end{minipage}
\end{table*}

\begin{figure*}
  \centering	
  \includegraphics[width=0.98\textwidth]{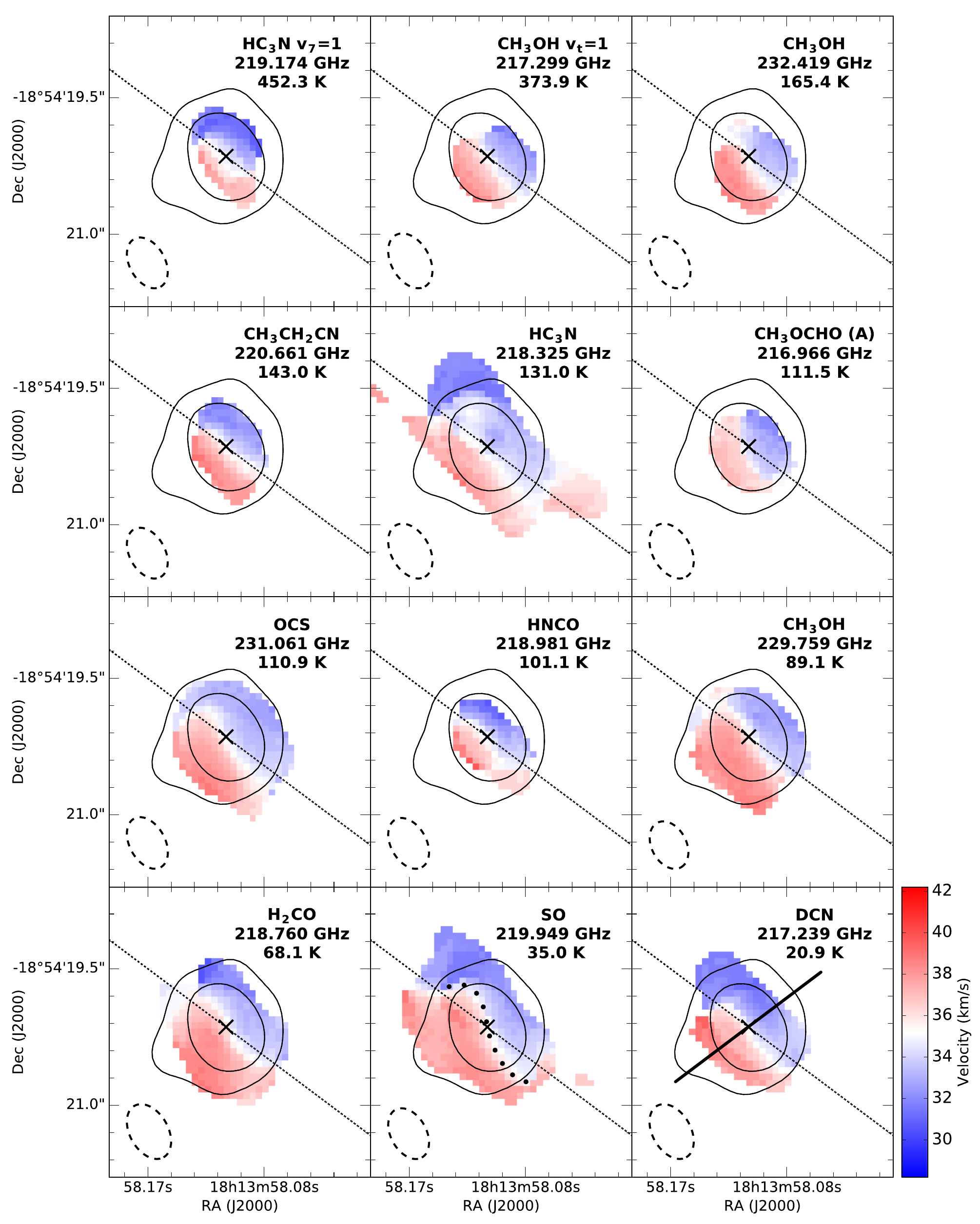}
  \caption{First moment maps (colourscale)  of selected spectral lines
    from Table~\ref{tab:linelist},  overlaid with contours of  the SMA
    VEX  1.3\,mm continuum  emission  (levels 5  \& 25$\sigma$,  where
    $\sigma=0.7$\,mJy beam$^{-1}$).  The colourscale is centred on the
    systemic    velocity    of    the    system,    35.2\,km\,s$^{-1}$
    \citep{cyganowski_2011sma}.   Each  panel   is  labeled  with  the
    species name  and the excitation  energy (in Kelvin) of  the upper
    level of the  transition.  The position angle of  the outflow from
    Fig~\ref{fig:outflow} is  shown with  a dashed black  line.  Black
    dots  in  the   lower  middle  panel  denote  the   twist  in  the
    zero-velocity  gas (Section~\ref{sec:results_compactlines}).   The
    cut used  to generate  the PV  diagrams in  Figure~\ref{fig:pv} is
    shown in the  lower right panel with a solid  black line. The beam
    is shown in the lower left of each panel as a dashed ellipse.}
    \label{fig:mom1_grid}
\end{figure*}

\begin{figure*}
  \centering	
  \includegraphics[width=0.98\textwidth]{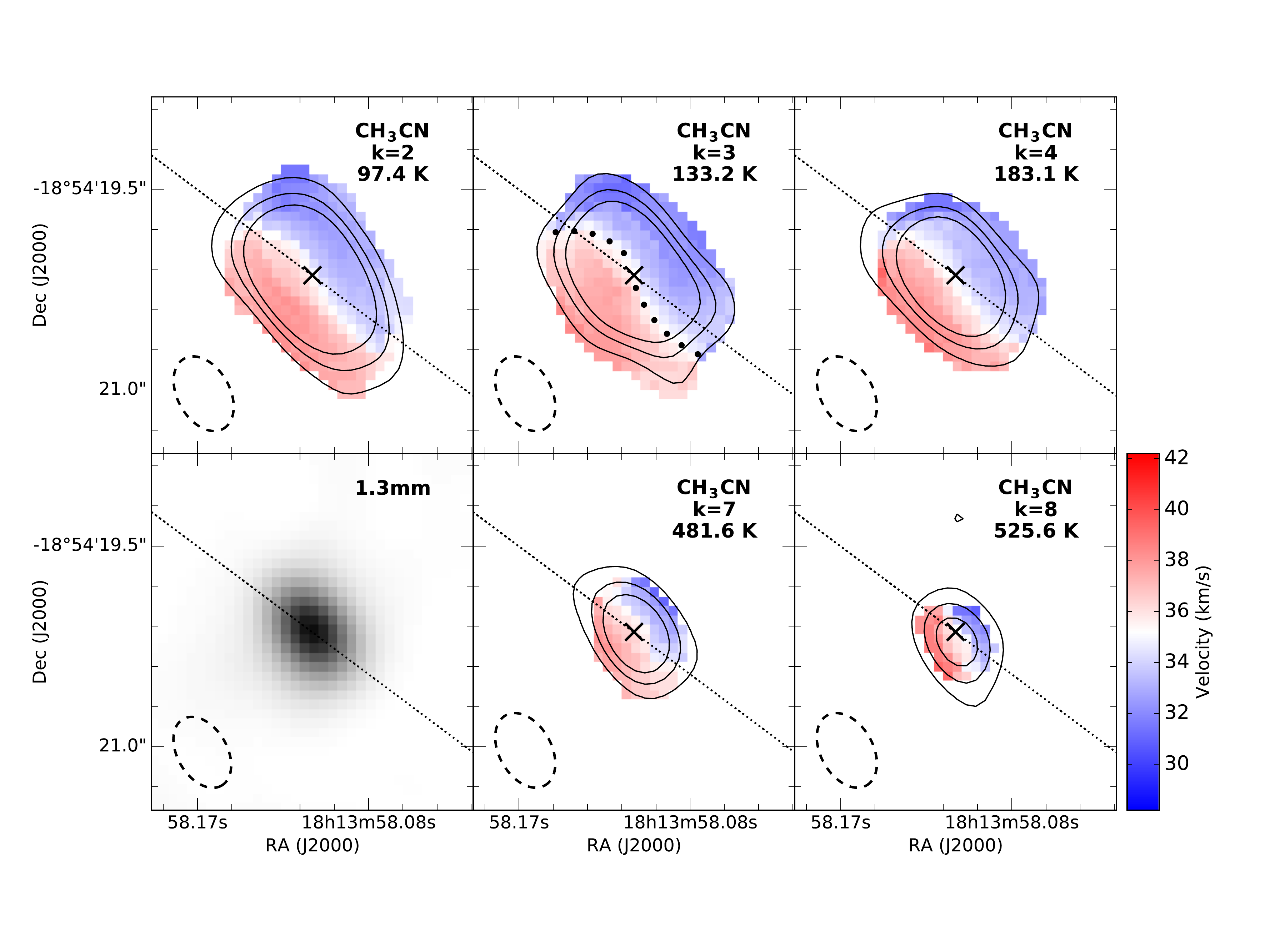}
  \caption{First   moment   maps  for   minimally-blended   CH$_{3}$CN
    transitions (we omit $K=6$ as it is blended with HNCO and $K=5$ as
    it is  blended with  CH$_{3}^{13}$CN $K=0$), overlaid  with zeroth
    moment  contours  and displayed  alongside  the  1.3 mm  continuum
    emission from Figure \ref{fig:outflow}.  The excitation energy (in
    Kelvin) of the upper level of each transition is indicated.  Black
    dots  in  the   upper  middle  panel  denote  the   twist  in  the
    zero-velocity  gas (Section~\ref{sec:results_compactlines}).   The
    continuum peak is marked with a black cross, and the beam is shown
    in  the lower  left  corner of  each panel  as  a dashed  ellipse.
    Contours are 3,  6 and 9\,$\sigma$, where $\sigma$  for the zeroth
    moment     maps    is     0.19,    0.22,     0.19,    0.16     and
    0.16\,$\mathrm{Jy}\,\mathrm{beam}^{-1}$, respectively.}
    \label{fig:ch3cn_mom1}
\end{figure*}

\smallskip

\begin{figure*}
  \centering
  \includegraphics[width=0.92\textwidth]{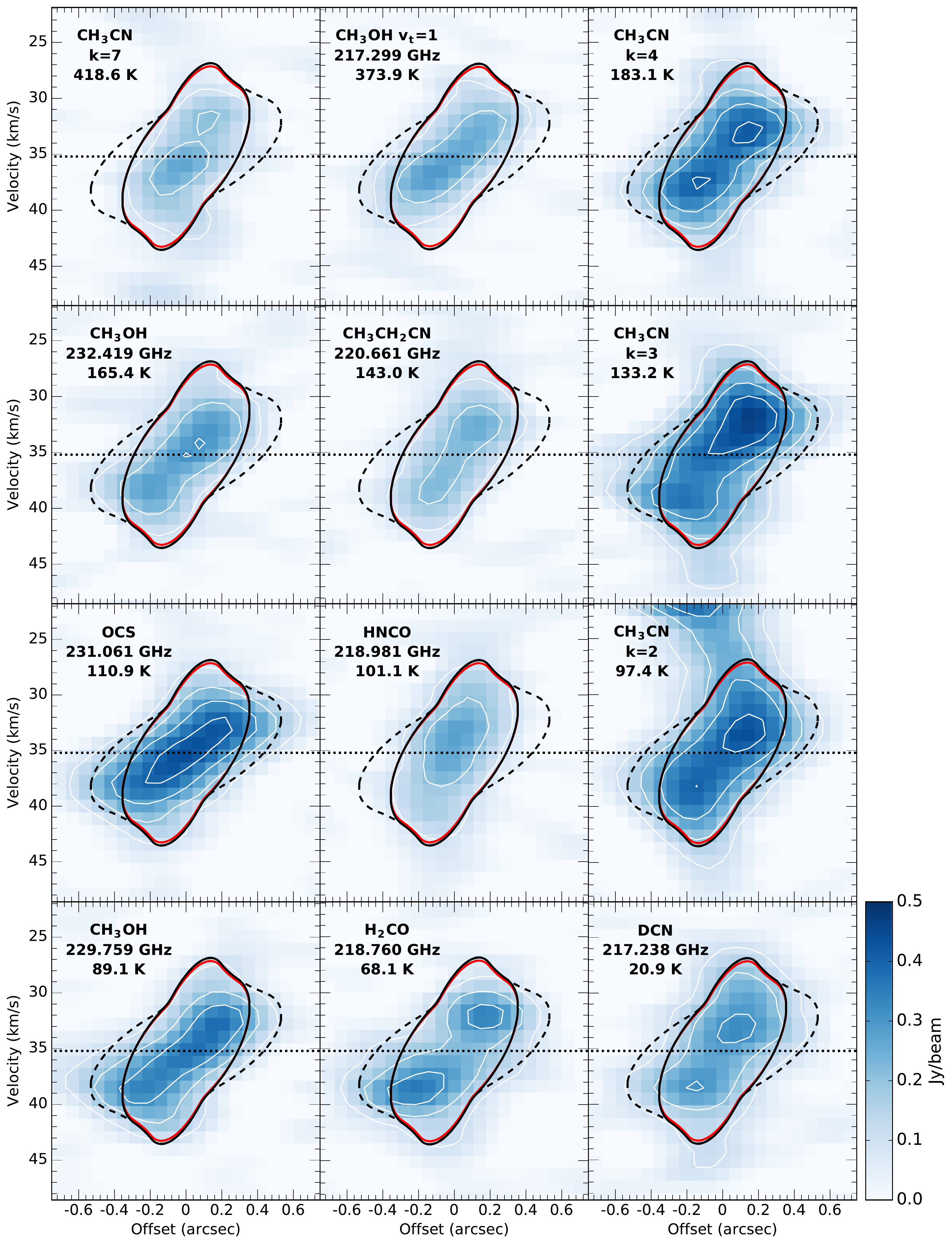}
  \caption{Position-velocity  diagrams  of selected  lines  from
      Table     \ref{tab:linelist},      ordered     by     decreasing
      E$_{\mathrm{upper}}$.  Solid white contours correspond to levels
      of    0.2,    0.3,    0.4   and    0.5\,Jy\,beam$^{-1}$.     The
      v$_{\mathrm{lsr}}$        of        the       system        from
      \citet{cyganowski_2011sma}, 35.2\,km\,s$^{-1}$, is  shown with a
      horizontal dotted  line; the  position of  the 1.3  mm continuum
      peak   from   the   SMA   VEX   observations   is   defined   as
      offset$=$0\farcs0.   The  CH$_{3}$CN $K=2$  transition  displays
      contamination at high blueshifted velocities from a nearby line.
      Overlaid    are   the    best   fitting    disc   models    from
      Section\,\ref{sec:kinematics}: the best fit model for all lines,
      excluding CH$_{3}$OH (M$_{\mathrm{enc}}$ = 60\,M$_{\odot}$, $i =
      35\degr$,  R$_{\mathrm{o}}$ =  1200\,au) is  shown with  a solid
      black line,  and the  best fit model  for CH$_{3}$OH  lines only
      (M$_{\mathrm{enc}}$   =   34\,M$_{\odot}$,    $i   =   52\degr$,
      R$_{\mathrm{o}}$ =  1200\,au) is  shown with  a solid  red line.
      These  two  models  are almost  identical  in  position-velocity
      space.  The dashed black line shows a model identical to that of
      the   best  fitting   disc  model   for  all   lines  (excluding
      CH$_{3}$OH), but  with R$_{\mathrm{o}}$ = 1800\,au,  providing a
      better fit to some transitions.}
  \label{fig:pv}
\end{figure*}

Figures~\ref{fig:mom1_grid} and \ref{fig:ch3cn_mom1}  present moment 1
maps  for  selected  lines from  Table~\ref{tab:linelist},  chosen  to
represent the  range of chemical species,  line emission morphologies,
kinematics, and  line excitation temperatures  present in our  SMA VEX
1.3 mm data.   Strikingly, all ten species show  a consistent velocity
gradient across the MM1 (sub)millimetre continuum source, with a sharp
transition from redshifted emission (to the South-East) to blueshifted
emission  (to  the  North-West).   The  orientation  of  the  velocity
gradient is consistent with that  seen at $\sim$2\farcs4 resolution by
\citet{cyganowski_2011sma} in  SO, HNCO,  CH$_3$OH, and  CH$_3$CN.  We
measured the position  angle of the velocity gradient for  each of the
31 transitions in Table~\ref{tab:linelist} by calculating the position
angle  of  the line  joining  the  RA/Dec  centroid positions  of  the
emission  in the  most redshifted  and the  most blueshifted  velocity
channels \citep[a  similar approach  to that  of][]{hunter_2014}.  The
median and standard deviation of the position angle across the fifteen
transitions that  we select as  potentially disc-tracing based  on our
kinematic fitting  and moment  analysis (Section~\ref{sec:kinematics},
Table~\ref{tab:linelist})  is  127$\pm$18$\degr$.   We note  that  the
variation  in the  median position  angle from  including a  different
selection  of lines  in the  calculation  is well  within the  scatter
indicated by the standard deviation of 18$\degr$, and adopt a position
angle  of  127$\degr$  for  constructing  the  position-velocity  (PV)
diagrams presented in Figure~\ref{fig:pv}.

\smallskip

Notably,  the  position  angle  of the  velocity  gradient  is  nearly
identical to  that of the  2D Gaussian model  fit to the  1.3\,mm dust
emission of  125$\pm$7$\degr$ \citep{cyganowski_2014}.   This alignment,
and the sharp transition from  redshifted to blueshifted emission, are
reminiscent  of   Keplerian  discs  observed  around   low-mass  stars
\citep[e.g.][]{hughes_2011,walsh_2014}.  The position  angles of MM1's
velocity gradient and  dust emission are also  nearly perpendicular to
that of  the high-velocity  bipolar molecular  outflow driven  by MM1.
The outflow  position angle,  estimated as the  position angle  of the
line joining  the peaks  of the  redshifted and  blueshifted $^{12}$CO
(3--2) lobes,  is $\sim$52$\degr$ (Figure~\ref{fig:outflow}).   For the
innermost  part  of   the  outflow  (shown  in  the   right  panel  of
Figure~\ref{fig:outflow}),   the    estimated   position    angle   is
$\sim$45$\degr$,  an offset  of 82$\degr$  from the  velocity gradient
seen in  the compact molecular  line emission.  Both estimates  of the
outflow  position   angle  are  consistent  with   the  outflow  being
perpendicular to the velocity gradient  in the compact gas, within the
scatter in  our estimate of  the latter.  This  configuration strongly
suggests  a  disc-outflow  system,   and  motivates  a  more  detailed
examination  of  the  kinematics  of  MM1  by  means  of  PV  diagrams
(Figure~\ref{fig:pv}).

\smallskip

As illustrated in Figure~\ref{fig:pv}, the MM1 PV diagrams exhibit the
characteristic pattern expected  for a Keplerian disc  --- the highest
velocities  are seen  closest to  the central  source, while  the most
spatially extended emission  is seen at lower  velocities (e.g. closer
to  the v$_{\mathrm{LSR}}$  of the  system).  These  key features  are
consistent  across  a  range   of  molecular  tracers  and  transition
excitation energies, and  support our use of Keplerian  disc models to
interpret the kinematics of MM1 (Section~\ref{sec:kinematics}).  There
are also,  however, some potentially illuminating  differences amongst
the prospective disc tracers.  Notably, most  of the lines do not peak
towards the central source, instead showing offsets in position and/or
velocity (Figure~\ref{fig:pv}:  the position of the  1.3\,mm continuum
peak from  our SMA  VEX observations  is defined  as offset=0\farcs0).
More surprisingly, these offsets do  not appear to be simple functions
of molecular abundance or transition excitation energy, in contrast to
the pattern  seen in  the candidate massive  disc NGC  6334I(N)-SMA 1b
\citep{hunter_2014}.  In MM1,  low-temperature transitions of abundant
molecules,             including             DCN(3--2)             and
H$_2$CO(3$_{\mathrm{2,1}}$--2$_{\mathrm{2,0}}$), show  a double-peaked
structure with a local minimum  towards the central source, consistent
with  a  radial temperature  gradient  (increasing  inwards, see  also
Section~\ref{sec:results_extent})  and  a moderately  edge-on  viewing
angle.   However,  higher-temperature  transitions of  CH$_{3}$CN  and
CH$_3$OH also show notably  double-peaked or asymmetric structure.  In
contrast,  the  peak of  the  OCS(19--18)  emission (with  a  moderate
E$_{\mathrm{upper}}$  of  110.9\,K)  is coincident  with  the  central
source.   Interestingly, CH$_{3}$CN  emission in  two other  candidate
discs   around   (proto)O-stars   \citep[CH$_{3}$CN   $K=3$   in   NGC
  6334I(N)-SMA    1b    and    $K=2$,     $K=4$,    and    $K=6$    in
  AFGL4176-mm1:][respectively]{hunter_2014,johnston_2015}     exhibits
asymmetries similar  to those  we observe  in CH$_{3}$CN  towards MM1,
though  these  asymmetries  are   not  reproduced  by  current  models
\citep[e.g.   Figure  4  of][]{johnston_2015}.   In  MM1,  as  in  NGC
6334I(N)-SMA  1b, HNCO  differs  from other  species  in displaying  a
compact morphology in  PV diagrams.  Based on the present  data, it is
difficult   to  disentangle   the  effects   of  molecular   abundance
\citep[e.g. potential  chemical segregation,  as observed  in AFGL2591
  by][]{jimenez-serra_2012} from those of  molecular excitation in the
MM1  disc.   Higher  angular resolution  (sub)millimetre  observations
would be  required to  obtain well-resolved  images of  many different
molecular species.

\smallskip

The  moment maps  shown in  Figure~\ref{fig:mom1_grid} also  highlight
that though  all species  share a  consistent velocity  gradient (from
redshifted  in  the  South-East  to blueshifted  in  the  North-West),
SO(6$_{\mathrm{5}}$--5$_{\mathrm{4}}$)  and   HC$_3$N  $v=0$  (24--23)
exhibit emission  that is notably  spatially extended compared  to the
millimetre  continuum  and  to  other   molecules.   The  SO  line  is
low-temperature (E$_{\mathrm{upper}}=35.0$\,K) and is clearly detected
in  the   G11.92--0.61  outflow  in  our   lower-resolution  SMA  data
\citep[e.g.  Figure  3 of][]{cyganowski_2011sma},  suggesting possible
outflow contribution  to the emission detected  in our high-resolution
SMA  VEX  observations.   The  extended emission  from  HC$_3$N  $v=0$
(24--23),      with       its      higher       excitation      energy
(E$_{\mathrm{upper}}=131.0$\,K),  is perhaps  more surprising,  though
recent  observations have  found that  HC$_3$N emission  is associated
with  gas   shocked  by  outflows  driven   by  low-mass  (proto)stars
\citep[e.g.][]{shimajiri_2015,benedettini_2013}.                Within
$\sim$10\,km\,s$^{-1}$ of the systemic velocity (e.g., the range shown
in  Figure~\ref{fig:mom1_grid}),  the  kinematics  of  outflow-tracing
molecules such as $^{12}$CO and HCO$^{+}$ are very confused, with both
redshifted and blueshifted emission detected to  both the NE and SW of
MM1  \citep[e.g.  Figures  7 and  8 of][]{cyganowski_2011sma}.   It is
thus plausible that  (the base of) the MM1 outflow  contributes to the
SO  and HC$_3$N  emission detected  in our  SMA VEX  observations.  In
addition,  given the  apparent elongation  along the  outflow axis  in
moment 0 maps of many of the  lines examined, it is also possible that
the base of the outflow contributes to more lines, though we note that
as  the beam  is  elongated in  a similar  direction,  this effect  is
difficult to quantify with our current observations.

\smallskip

Several of the  moment 1 maps (e.g.   CH$_3$CN $K=2,3,4$, OCS(19--18),
H$_{2}$CO(3$_{\mathrm{2,1}}$--2$_{\mathrm{2,0}}$),    SO($6_{5}$    --
$5_{4}$),  OCS(19--18) and  DCN(3--2)) also  exhibit an  interestingly
asymmetric feature: a  twisted structure that is most  clearly seen by
examining the  gas moving at  the systemic velocity (white  in Figures
\ref{fig:mom1_grid} \& \ref{fig:ch3cn_mom1},  marked with black dots).
Based  on  the  orientation  of the  blue-  and  red-shifted  emission
(i.e. in  opposition to the  movement of the outflowing  material), we
suggest that  this twisting is likely  due to the effect  of infalling
material.   Such  an  interpretation   is  also  strengthened  by  the
requirement of an infall component to  be used in our modelling of the
PV diagrams (see Figure \ref{fig:pv} and Section \ref{sec:kinematics})

\subsection{Extent of gas and dust emission in the disc}
\label{sec:results_extent}

The thermal  dust emission from MM1  is unresolved in our  SMA 1.3\,mm
VEX   observations,  with   a  fitted   deconvolved  source   size  of
$1150\times960$\,au   \citep{cyganowski_2014}.    In   contrast,   the
molecular line emission  from the disc is marginally  resolved in some
lower-excitation transitions.  For example,  for the blended $K=0$ and
$K=1$ components of  CH$_3$CN(12--11), the deconvolved size  of the 2D
Gaussian      fit      to      the      moment      0      map      is
0\farcs80$\pm$0\farcs06$\,\times\,$0\farcs53$\pm$0\farcs04
($\mathrm{P.A.}=47\pm8$),  or $\sim2700\times1800$\,au.   The CH$_3$CN
ladder is of  particular interest, as the relative  spatial extents of
the different  k-components are  expected to  depend primarily  on gas
temperature.    Figure~\ref{fig:ch3cn_mom1}   presents   contours   of
integrated intensity  for the  $K=2$, $K=3$,  $K=4$, $K=7$,  and $K=8$
transitions of CH$_3$CN(12--11), which range in excitation energy from
E$_{\mathrm{upper}}=97.4$ --  525.6\,K, overlaid on the  moment 1 maps
of these transitions (we omit $K=0$ and $K=1$ because they are blended
with    each   other,    $K=6$    because   it    is   blended    with
HNCO(10$_{1,9}$-9$_{1,8}$ at  220.585 \,GHz,  and $K=5$ because  it is
blended    with   CH$_{3}^{13}$CN    K=$0$).    As    illustrated   by
Figure~\ref{fig:ch3cn_mom1},   all  transitions   display  a   similar
velocity  gradient, but  the  low-$k$,  low-excitation transitions  of
CH$_3$CN   are  significantly   more  extended   than  the   high-$k$,
high-excitation  transitions.   This  pattern  of  decreasing  spatial
extent  with   increasing  excitation  energy  is   suggestive  of  an
increasing  temperature  gradient  towards   a  central  source.   The
temperature  structure  of the  gas  around  MM1, as  determined  from
modelling   the   CH$_3$CN   emission,   is   discussed   further   in
Section~\ref{dis:cassis}.

\begin{figure}
  \includegraphics[width=\columnwidth]{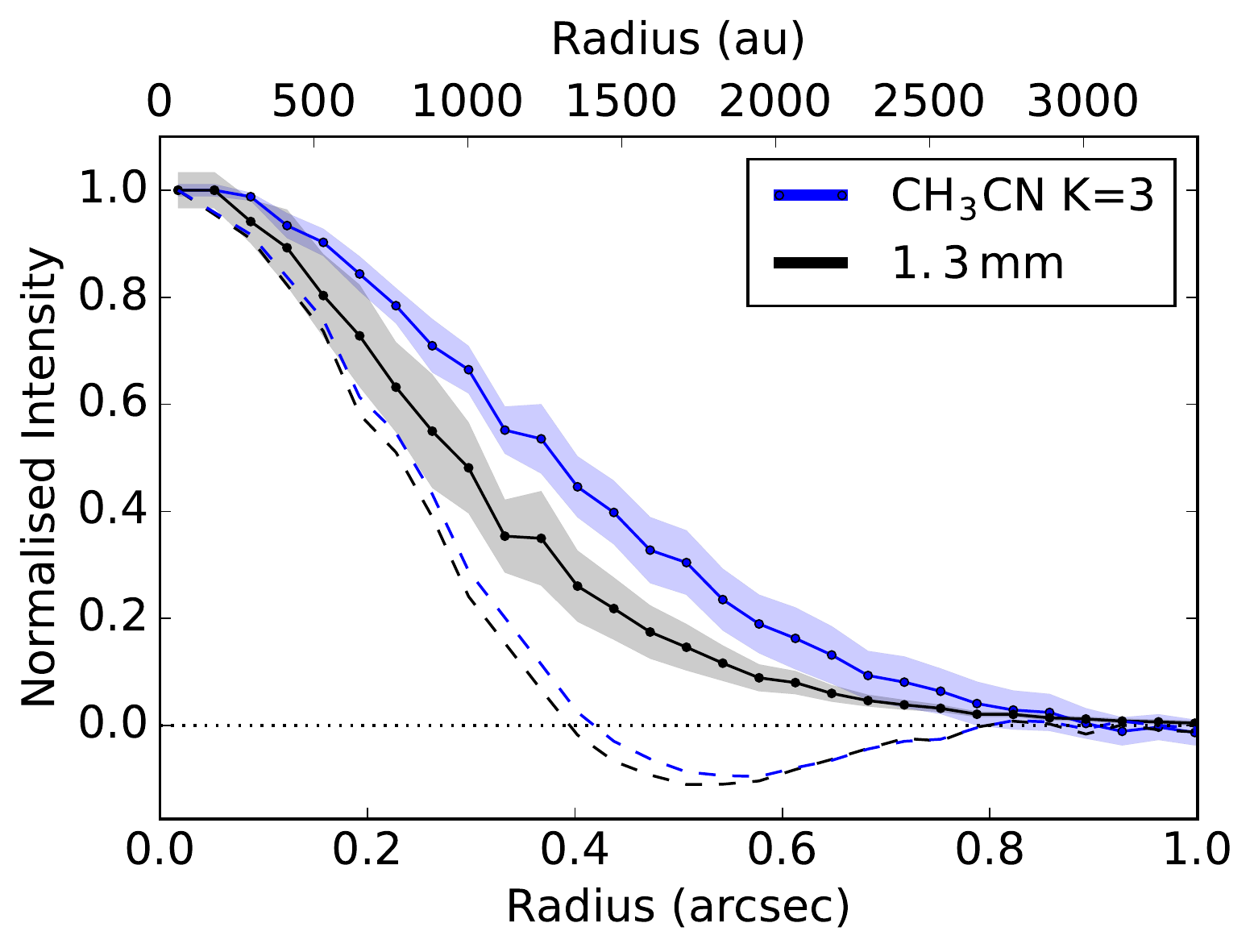}
  \caption{Azimuthal  averages  of  the normalised  intensity  of  the
    CH$_{3}$CN  $K=3$  line  (blue)  and  1.3  mm  continuum  emission
    (black), centered around  the bi-section between the  most red and
    blue  shifted channel  centroids, with  1$\sigma$ errors  shown as
    shaded regions.   Azimuthal averages of  the beams are  shown with
    dashed lines.}
  \label{fig:azimuthal}
\end{figure}

\smallskip

To investigate the differing extents of  the dust and gas emission, we
performed  simple azimuthal  averages  (i.e.\ not  accounting for  any
ellipicity effects) on both the  continuum and the CH$_3$CN $K=3$ line
emission (the  lowest-energy k component  that is well  separated from
neighboring      lines,      see      also      top      panel      of
Figure~\ref{fig:ch3cn_cassis}).       As      shown     in      Figure
\ref{fig:azimuthal}, this analysis confirms  that the CH$_{3}$CN $K=3$
emission is more extended than  the 1.3 mm continuum, by approximately
0\farcs1  at  maximum   deviation.   Figure  \ref{fig:azimuthal}  also
highlights  that  the  continuum   emission,  unlike  the  $K=3$  line
emission, is  unresolved by  our SMA  observations.  A  larger spatial
extent of gas  emission in comparison to dust continuum  emission is a
result  often  seen  in  observations of  circumstellar  discs  around
lower-mass                         young                         stars
\citep[e.g.][]{guilloteau_2011,perez_2012,de-gregorio-monsalvo_2013}.
In such systems,  the spatial difference is often suggested  to be due
to either  radial migration of the  dust towards the central  star, or
viscous outward spreading of the gas \citep[e.g.][]{alexander_2006}.

%%%%%%%%%%%%%%%%%%%%%%%%%%%%%%%%%%%%%%%%%%%%%%%%%%%%%%%%%%%%%%%%%%%%%%
\section{Discussion}                                                 %
\label{sec:discussion}                                               %
%%%%%%%%%%%%%%%%%%%%%%%%%%%%%%%%%%%%%%%%%%%%%%%%%%%%%%%%%%%%%%%%%%%%%%

\subsection{Molecular line modelling: gas kinematics}
\label{sec:kinematics}

\begin{figure*}
    \flushleft
   \includegraphics[width=0.901\textwidth]{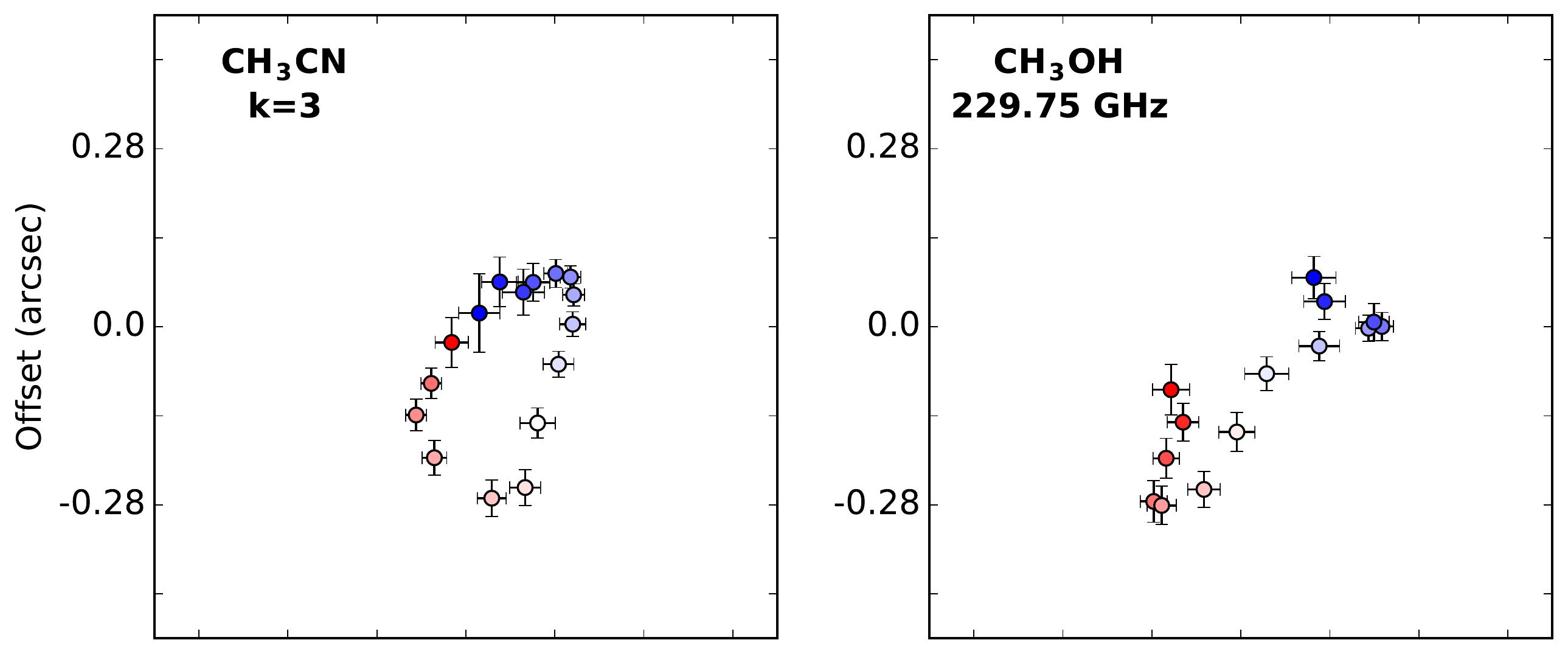}
   \includegraphics[width=\textwidth]{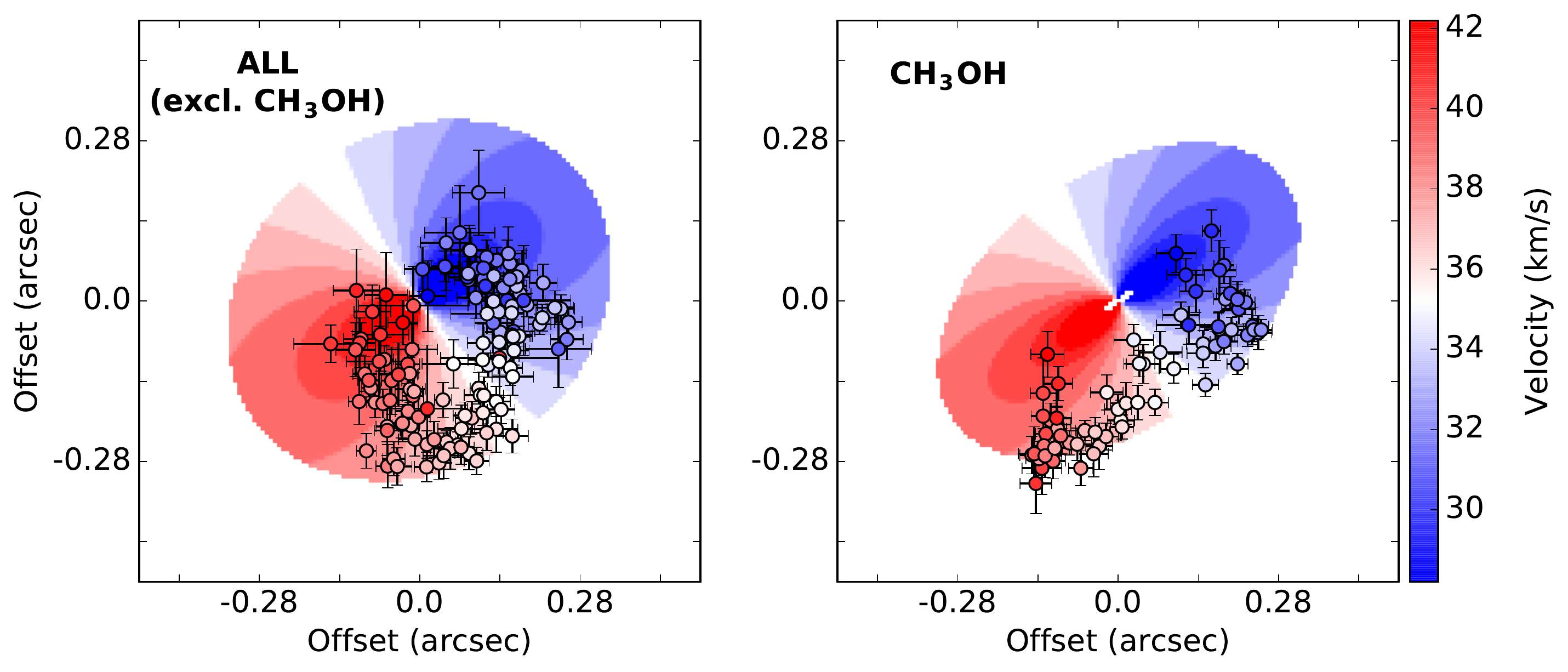}
   \caption{Top: centroid plots for the CH$_{3}$CN K=3 (left) and
       the CH$_{3}$OH 229.75\,GHz  (right) transitions, displaying the
       differing  morphology  characteristic of  CH$_{3}$OH.   Bottom:
       Results  of  the  kinematic   fitting  procedure  described  in
       Section~\ref{sec:kinematics}   for  the   two  collections   of
       centroid  points   ---  all   'disc-tracing'  lines   other  than
       CH$_{3}$OH  (left)  and 'disc-tracing'  CH$_{3}$OH  transitions
       only  (right).   The best  fitting  disc  model for  all  lines
       (except   CH$_{3}$OH)   surrounded    an   enclosed   mass   of
       $60^{+21}_{-27}$\,M$_{\odot}$    at     an    inclination    of
       $35^{+20}_{-6}\degr$,  with an  outer radius  R$_{\mathrm{o}} =
       1200^{+100}_{-100}$\,au.  The best fitting  disc model for only
       the  CH$_{3}$OH  transitions  surrounded an  enclosed  mass  of
       $34^{+28}_{-12}$\,M$_{\odot}$    at     an    inclination    of
       $52^{+11}_{-14}\degr$, with  an outer radius  R$_{\mathrm{o}} =
       1200^{+100}_{-100}$\,au.  In both cases,  the position angle of
       the disc  was fixed  to 127$\degr$.  Axes  are in  offsets with
       respect    to   the    velocity    centre    as   defined    in
       Section~\ref{sec:kinematics}.}
  \label{fig:all_centroids}
\end{figure*}

In order to further investigate  the kinematic origin of each emission
line, we  fitted a  2-dimensional Gaussian to  each channel  using the
\textsc{casa imfit} routine.   For a given line and  channel, this fit
provides the  location of the centroid,  whose position on the  sky is
recorded  along  with the  channel  velocity.   This process  is  then
repeated for all channels across a  given line.  The morphology of the
resulting centroid plots often exhibits  a closed loop structure, with
the  most  red-  and   blue-shifted  channels  closest  together,  and
velocities between these  channels tracing an arc across  the sky 
  (Figure~\ref{fig:all_centroids}).  However,  the size of  this loop
varies  from line  to line,  with  some lines  exhibiting more  linear
centroid plots (e.g.  the CH$_{3}$OH  transitions).  For each line, we
define  a velocity  centre  ---  the on-sky  position  that bisects  the
positions of  the most red-  and blue-shifted channel  centroids.  The
position  of this  velocity centre  for each  line differed  slightly,
however  all were  consistent  to within  approximately 0\farcs05  (or
$\sim200$\,au at the distance of MM1).

\smallskip

Based on careful examination of the individual line centroid plots and
the corresponding moment  1 maps, we selected a  subset of potentially
disc-tracing (DT)  lines.  In order  to model the kinematic  origin of
these  lines, we  adopted an  approach  similar to  that presented  in
\citet{sanchez_2013},  and  use  a   model  of  a  geometrically  thin
Keplerian  disc  \citep{maret_2015}  to   compare  with  the  centroid
information.   The  model disc  is  characterised  by a  central  mass
(M$_{\star}$), an  outer radius (R$_{\mathrm{o}}$) and  an inclination
to the line of sight ($i$, where $i=0$ corresponds to a face on disc),
and is  assumed to emit uniformly  as a function of  radius (though we
note that  adopting different  intensity distributions does  not alter
the results of our fitting).  The  model is given a number of velocity
channels with width  consistent with that of our  observations, and is
assumed to lie at the same distance as MM1 (3.37\,kpc).

\smallskip

For each  combination of free  parameters, a spectral cube  is created
for the model, and then a first  moment map is produced.  This is then
compared to the centroid plots.  The chi-squared landscape is explored
in  a  grid based  fashion  using  the \textsc{scipy  optimize  brute}
module,   where   the   ranges   for  each   parameter   were   $1   <
\mathrm{M}_{\star} < 90$\,M$_{\odot}$,  $600 < \mathrm{R}_{\mathrm{o}}
< 1400$\,au, and $0 < i < 90\degr$.  Once a minimum within the grid is
recovered, a  Nelder-Mead (or \textsc{ameoba})  minimisation technique
is used  to refine the best  fitting parameters.  Due to  the slightly
different morphologies of  the CH$_{3}$OH centroid plots,  we chose to
perform the fitting  on two collections of centroid  locations --- those
obtained from  all lines but excluding  CH$_{3}$OH (labelled `DT\_ALL'
in Table\,\ref{tab:linelist}), and only the CH$_{3}$OH lines (labelled
`DT\_CH$_{3}$OH' in Table\,\ref{tab:linelist}).  The best fitting disc
model    for   all    lines    surrounds   an    enclosed   mass    of
$60^{+21}_{-27}$\,M$_{\odot}$      at      an      inclination      of
$35^{+20}_{-6}\degr$.   The  best  fitting  disc model  for  only  the
CH$_{3}$OH    transitions    surrounds    an    enclosed    mass    of
$34^{+28}_{-12}$\,M$_{\odot}$      at      an      inclination      of
$52^{+11}_{-14}\degr$. In  both cases, the disc  extended to 1200\,au.
Figure \ref{fig:all_centroids} shows the resulting best fitting models
for each of these collections.

\smallskip

Based  on the  results of  the fitting  to the  centroid maps,  we can
follow   the   prescription   of  \cite{cesaroni_2011}   and   overlay
theoretical  position-velocity models  for  these  Keplerian discs  on
Figure \ref{fig:pv}.  In this case,  the region on a position-velocity
diagram within which emission is expected can be expressed as
\begin{equation}
V = \sqrt{GM}\frac{x}{R^{\frac{3}{2}}} + \sqrt{2GM}\frac{z}{R^{\frac{3}{2}}}, 
\end{equation}
where the first term is the  contribution from the Keplerian disc, and
the second  the contribution due  to free  fall.  $V$ is  the velocity
component along the  line of sight, $M$ is the  enclosed mass, $x$ and
$z$  are the  co-ordinates along  the disc  plane and  line of  sight,
respectively, and  $R =  \sqrt{x^2 +  z^2}$ is  the distance  from the
centre of the disc.

\smallskip

The  models overlaid  on Figure  \ref{fig:pv} correspond  to the  best
fitting solutions  found from  the centroid fitting  --- the  best fit
model   for   all   `disc-tracing'   lines   other   than   CH$_{3}$OH
(M$_{\mathrm{enc}}$ = 60\,M$_{\odot}$, $i = 35\degr$, R$_{\mathrm{o}}$
= 1200\,au,  based on nine  transitions) is  shown with a  solid black
line,  and the  best  fit model  for  `disc-tracing' CH$_{3}$OH  lines
(M$_{\mathrm{enc}}$ = 34\,M$_{\odot}$, $i = 52\degr$, R$_{\mathrm{o}}$
= 1200\,au, based on six transitions)  is shown with a solid red line.
These models produce almost identical projections in position-velocity
space.  For all transitions, the model accurately reproduces the range
of velocities displayed  in the data.  The major  differences arise in
the  spatial   extent  of  the   emission,  with  molecules   such  as
CH$_{3}$CH$_{2}$CN,  HNCO,  DCN  and  CH$_{3}$CN  $K=7$  appearing  to
originate from smaller  radii than molecules such as OCS  or the lower
$K$ transitions  of CH$_{3}$CN.  For  this reason, we also  overplot a
model  identical  to  the  best  fitting  disc  model  for  all  lines
(excluding  CH$_{3}$OH), but  with R$_{\mathrm{o}}  = 1800$\,au,  as a
dashed  black  line.   Interestingly,  while   there  seems  to  be  a
dependence of location/extent of emission on excitation energy for the
CH$_{3}$CN transitions, there appears not to be a similar trend across
all  molecules.   This  can  most  easily be  seen  by  comparing  DCN
(E$_{\mathrm{upper}}  =   20.9$\,K)  with,  for   example,  CH$_{3}$OH
$v_{t}=1$ (E$_{\mathrm{upper}}  = 373.9$\,K),  whose PV  diagrams show
similar spatial extents.

\smallskip

We can also estimate an inclination angle from the observed ellipicity
of the  deconvolved Gaussian model  for the SMA VEX  1.3\,mm continuum
emission  \citep{cyganowski_2014}.  Assuming  circular symmetry,  this
implies    an   inclination    angle    for   the    dust   disc    of
$\sim31^{+5\degr}_{-7\degr}$.   While we  interpret  this result  with
some caution because the continuum  is unresolved by our observations,
this  estimate  provides  an  independent   line  of  evidence  for  a
moderately inclined disc.

\subsection{Molecular line modelling: gas temperatures and physical properties from CH$_{3}$CN}
\label{dis:cassis}

To determine the physical properties of  the gas around MM1, we follow
the  approach of  \citet{hunter_2014} and  use the  CASSIS package  to
model  the  line  intensities  and  profiles  of  the  CH$_{3}$CN  and
CH$_{3}^{13}$CN  emission  line  ladders,  which have  been  shown  to
provide robust  measurements of gas  physical conditions in  hot cores
\citep[e.g.][]{pankonin_2001, araya_2005}.

\smallskip

In  order  to  quantify  any  spatial  variations  in  these  physical
conditions, we extract the CH$_3$CN  spectra on a pixel-by-pixel basis
across the location  of MM1.  The physical model used  to fit the data
was  assumed  to be  in  local  thermodynamic equilibrium  (LTE),  and
possessed six free parameters  --- CH$_{3}$CN column density, excitation
temperature, line  width, emission region diameter,  velocity, and the
isotopic ratio of  $^{12}$C / $^{13}$C.  The ranges  explored for each
parameter   were   as   follows   ---  column   density:   $10^{16}   <
N_{\mathrm{CH}_{3}\mathrm{CN}}  \leq  10^{18}$\,cm$^{-2}$;  excitation
temperature:  $40 <  T_{\mathrm{ex}} \leq  250$  K; line  width: $3  <
\Delta  \nu \leq  8$\,km\,s$^{-1}$;  size: $0.1\arcsec  < \theta  \leq
0.5\arcsec$; velocity:  $31 <  v < 42$\,km\,s$^{-1}$;  isotopic ratio:
$55 < \,^{12}\mathrm{C}  / ^{13}\mathrm{C} \leq 85$.   The fitting was
performed  using  a  Markov-Chain   Monte  Carlo  (MCMC)  minimisation
strategy  ---  an  initial  guess  for  each  parameter  is  taken,  and
randomised steps in each parameter  are taken to explore the resulting
goodness of fit.  These steps are initially large, but decrease as the
fitting  procedure progresses  in  order to  refine  the best  fitting
parameters.  The number of iterations was  set to 5000, and the cutoff
parameter to  determine when the  step size  becomes fixed was  set to
2500.   The best  fitting values  for each  pixel are  taken from  the
execution that  achieved an  acceptance rate of  0.5.  To  ensure only
data with  a reliable signal-to-noise  ratio were used in  the fitting
procedure, only  pixels with  spectra where  the $K=3$  transition was
detected  above  $5 \sigma$  were  included  in the  modelling,  which
ensured  the fitting  routine was  always given  at least  the $K=0/1$
(blended with  each other), $K=2$  and $K=3$ transitions to  fit.  The
subsequent best fitting  model spectra were then examined  by eye, and
any inadequate fits were discarded from the final results.

\smallskip

In all cases, we found that models with a single component of emitting
material could  not adequately reproduce  the line ratios  observed in
the CH$_{3}$CN  and CH$_{3}^{13}$CN spectra.  Instead,  two components
of  emitting  material  were  required  to  adequately  reproduce  the
emission    \citep[similar    to    observations   of    hot    cores,
  e.g.][]{cyganowski_2011sma, hernandez-hernandez_2014}.  In these two
component  fits,  each  component  was  given  a  set  of  independent
parameters as listed above.

\smallskip

\begin{figure*}
\includegraphics[height=0.9\textheight]{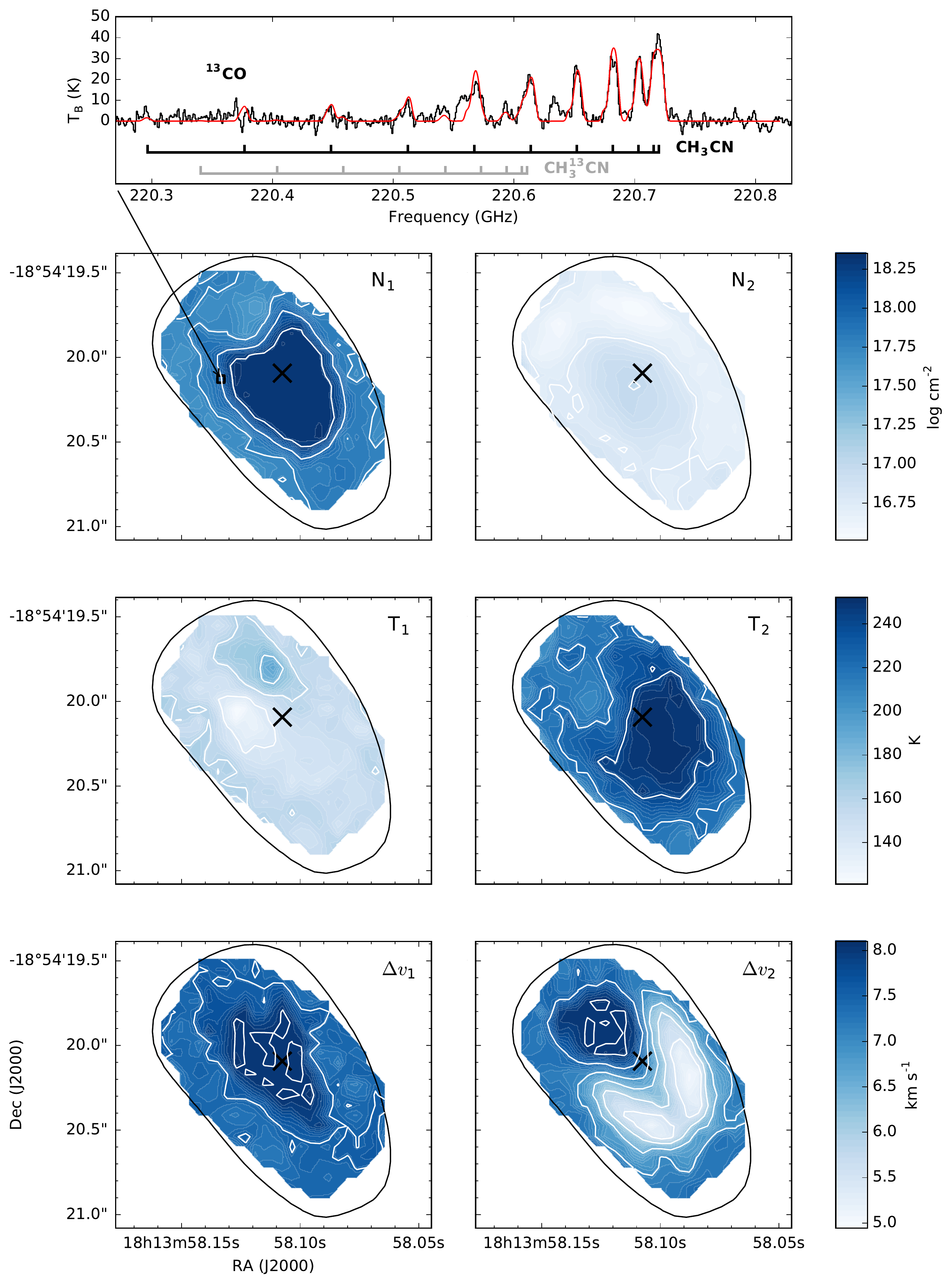}
  \caption{Results  from  CASSIS  fitting to  the  CH$_{3}$CN  spectra
    across the region of MM1.   \textit{Top:} fitting results (red) to
    the spectrum (black) from a single pixel.  \textit{Bottom:} Column
    densities ($N$),  temperatures ($T$) and line  widths ($\Delta v$)
    as derived from the fits for the first component (left) and second
    component  (right).   White  contours correspond  to  the  labeled
    values with tick marks on each colourbar.  The black contour shows
    the 3$\sigma$ level  of the moment 0 map for  the $K=2$ transition
    (where $\sigma=0.19$\,Jy  beam$^{-1}$), and the black  cross marks
    the location of the 1.3mm continuum peak.}
  \label{fig:ch3cn_cassis}
  \vspace{1cm}
\end{figure*}

Figure~\ref{fig:ch3cn_cassis}  shows   the  results  of   the  fitting
procedure.  The top panel shows the  results of a fit to an individual
pixel,   with   the   respective   transitions   of   CH$_{3}$CN   and
CH$_{3}^{13}$CN labelled, along with  the transition of $^{13}$CO that
appears within  that range  of frequencies.   There is  good agreement
between the  line ratios and  line widths of  the model and  the data,
suggesting we are  placing strong constraints on  the parameters.  The
lower panels of Figure~\ref{fig:ch3cn_cassis} show the results for the
first  component  (left)  and  second component  (right).   The  first
component is characterised by cooler  material ($\sim 150$\,K), with a
higher  column  ($\sim  10^{18}$\,cm$^{-2}$) and  a  larger  linewidth
($\sim 8$\,km\,s$^{-1}$), while the  second component is warmer ($\sim
250$\,K) with a lower  column ($\sim 10^{16}$\,cm$^{-2}$) and exhibits
a  range of  linewidths (5--8\,km\,s$^{-1}$).   These two  temperature
components  may   be  identified  with  two   distinct  reservoirs  of
CH$_{3}$CN  in  the  disc,  as  seen  in  recent  chemical  models  of
circumstellar  discs  around low  to  intermediate  mass young  stars.
These models show that substantial  reservoirs of CH$_{3}$CN can exist
in both the disc midplane  and disc atmosphere \citep{walsh_2015}.  In
the  upper  regions  of  these  disc  models,  ion-molecule  chemistry
dominates the production  of CH$_3$CN (with a  small contribution from
the thermal  desorption of  ice mantles), and  strong UV  fluxes (when
present) dominate the destruction  through photodissociation.  For the
midplane regions, which  hold the majority of  the CH$_3$CN reservoir,
production  is dominated  by thermal  desorption from  ice mantles  in
regions where the temperature exceeds  150\,K, and the molecule simply
freezes  out in  regions  with lower  temperatures (C.~Walsh,  private
communication).  It is  interesting to note that  while the parameters
of the first, lower-temperature  component appear relatively symmetric
about the continuum peak, the  temperature and linewidth of the second
component exhibit significant asymmetries  towards the South-West.  In
particular, the high temperatures seen towards the South-West would be
consistent   with   a   disc   oriented   as   modelled   in   Section
\ref{sec:kinematics}, if the disc was sufficiently flared and the disc
was  optically  thick.   Such  a   geometry  would  also  explain  the
orientation  of  the looped  centroids  of  emission seen  in  Section
\ref{sec:kinematics}.

\subsection{Nature of the central cm-wavelength source}
\label{sec:ff_sed}

\begin{figure}  % Fig 8
  \includegraphics[width=\columnwidth]{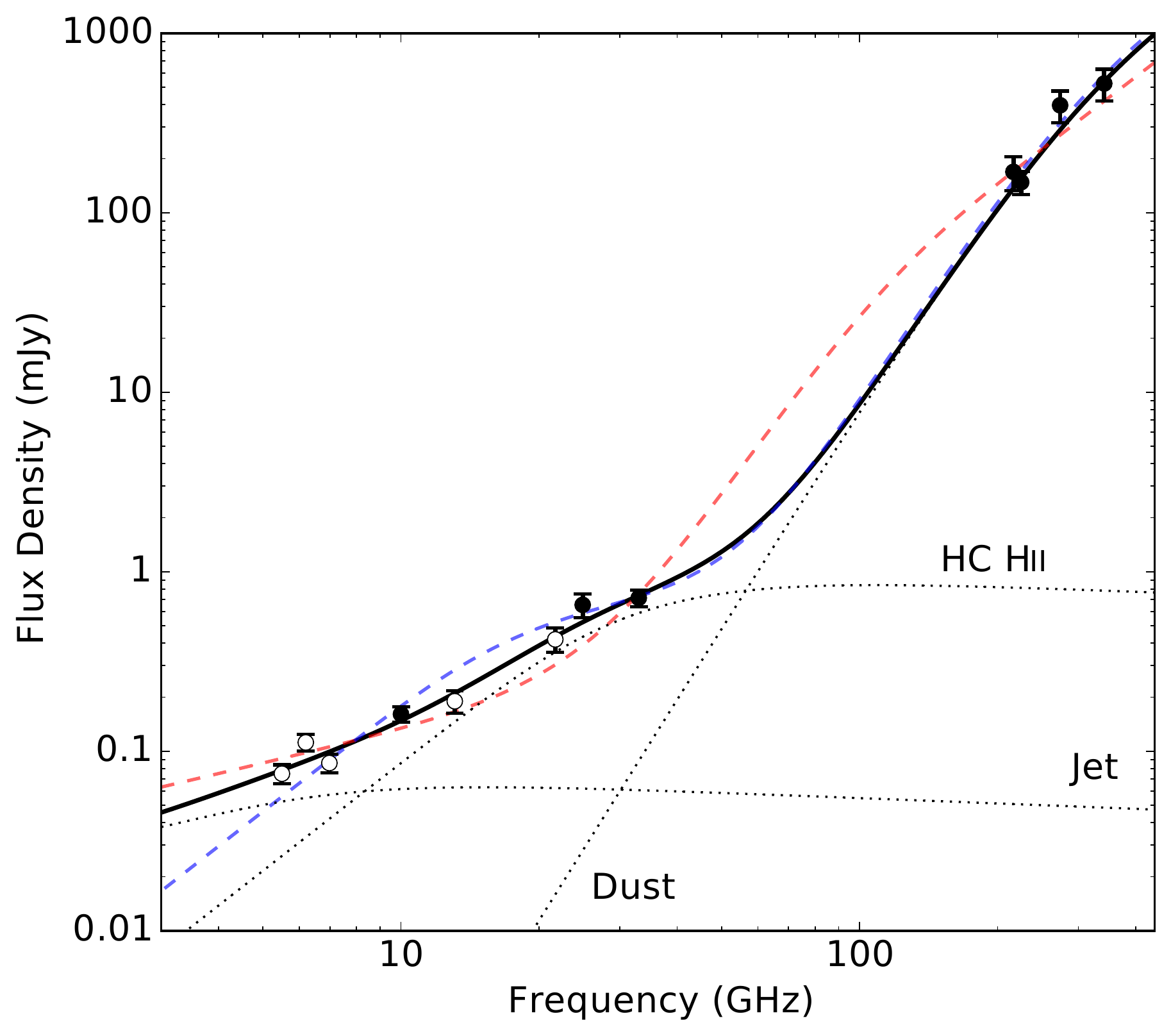}
  \caption{Spectral  energy distribution  of  MM1  from centimetre  to
    submillimetre wavelengths including data  from this paper (3.0 and
    0.9\,cm  VLA),  \citet{cyganowski_2014}   (1.2\,cm  VLA,  1.3\,mm,
    1.1\,mm,   and  0.88\,mm   SMA),  and   \citet{cyganowski_2011sma}
    (1.4\,mm CARMA),  shown as filled points.   These measurements are
    integrated flux  densities from two-dimensional  Gaussian fitting;
    the error  bars represent  the statistical uncertainties  from the
    fit  combined in  quadrature with  conservative estimates  for the
    absolute       flux       calibration       uncertainty       (see
    Section~\ref{sec:ff_sed}).    Shown  as   open  points   are  flux
    densities from  \citet{moscadelli_2016} (4.3-5.5, 2.3  and 1.4\,cm
    VLA, see Table~\ref{tab:sed}).  The best fitting model, shown as a
    solid  black  line,  combines  dust emission,  a  uniform  density
    HC~H\,{\sc ii} region, and an ionised jet with a power-law density
    profile (labelled black dotted lines).  For comparison, the dashed
    red line  and dashed blue  line show a dust  and jet model,  and a
    dust      and     HC~H\,{\sc      ii}     model,      respectively
    (Section~\ref{sec:ff_sed}).}
    \label{fig:sed}
\end{figure}

\begin{table}
  	\centering
	\begin{minipage}{0.9\columnwidth}
	\centering
	\caption{Measured continuum flux densities included in the spectral
          energy distribution shown in Figure~\ref{fig:sed}.}
	\label{tab:sed}
	\begin{tabular}{lccc}
         \hline
	Wavelength 	& Frequency	& Flux Density & Reference$^{a}$	 	\\
                   	&      (GHz)  		& (mJy) 	&	\\ 
        \hline
	5.5 cm & 5.50 		& $0.075 \pm 0.008$ & M16  \\
	4.8 cm & 6.2 		& $0.112 \pm 0.011$ & M16  \\
	4.3 cm & 6.98 		& $0.086 \pm 0.009$ & M16  \\
	3.0 cm & 10.0  		& $0.161 \pm 0.014$ & This work	\\
	2.3 cm & 13.1 		& $0.19  \pm 0.02$ & M16   \\
	1.4 cm & 21.7 		& $0.42  \pm 0.05$ & M16 	\\
	1.2 cm & 24.9 		& $0.654 \pm 0.082$ & C14  \\
	0.9 cm & 33.0 		& $0.715 \pm 0.023$ & This work  \\
	1.4 mm & 216.5 		& $169   \pm 13$ & C11	\\
	1.3 mm & 225.1 		& $148   \pm 3$ & C14 	\\
	1.1 mm & 273.7 		& $397   \pm 6$ & C14  	\\
	0.88 mm & 341.6 		& $525   \pm 8$ &C14 	\\
        \hline
	\end{tabular}
	\begin{flushleft}
          \small{$a$: M16: \citet{moscadelli_2016}; C14:
            \citet{cyganowski_2014}; C11: \citet{cyganowski_2011sma}}\\
	\end{flushleft}
\end{minipage}
\end{table}

In  order to  constrain the  properties of  the central  cm-wavelength
source, we model the free-free and dust emission simultaneously, using
an  approach  similar  to   that  of  \citet{hunter_2014}.   We  first
construct  the spectral  energy  distribution (SED)  of  MM1 shown  in
Figure~\ref{fig:sed} (filled  points) from  our new  VLA observations,
the  four  datapoints  in \citet{cyganowski_2014}  (1.2\,cm,  1.3\,mm,
1.1\,mm,  and  0.88\,mm),  and  the  1.4\,mm  CARMA  measurement  from
\citet{cyganowski_2011sma}.  We choose these measurements because they
are, by  design, as comparable  as possible in angular  resolution and
\emph{uv}-coverage (the  new 3.0  and 0.9\,cm VLA  data are  of higher
resolution than the other datasets, but designed to be well-matched to
each other).  The error bars plotted in Figure~\ref{fig:sed} represent
conservative estimates  of the  absolute flux  calibration uncertainty
(15 per cent for SMA 1.3\,mm VEX, 5 per cent for VLA $\lambda>2.5$~cm,
10 per cent for VLA $\lambda<2.5$~cm, 20 per cent for higher-frequency
SMA and for  CARMA 1.4\,mm), added in quadrature  with the statistical
uncertainties from the Gaussian fitting.

\smallskip

To  model the  centimetre-submillimetre wavelength  SED, we  combine a
free-free model \citep{olnon_1975}  with a single-temperature modified
graybody     function      representing     the      dust     emission
\citep[e.g.][]{gordon_1995,  rathborne_2010}.   We explored  modelling
the  free-free component  with  the  various \citet{olnon_1975}  model
geometries for the ionised gas: spherical (uniform density), Gaussian,
power  law,   and  truncated   power  law  ($n_{e}   \propto  r^{-2}$,
transitioning  to  a  central   constant-density  sphere  to  avoid  a
singularity).  We  found that  the uniform  density sphere  model best
reproduces  the centimetre-wavelength  shape  of the  SED ($\lambda  =
0.9-3$~cm), so we focus on that  model for our detailed exploration of
parameter  space.    The  initial   combined  model  has   seven  free
parameters:  the electron  density  (n$_{e}$),  radius (R$_{e}$),  and
electron temperature (T$_{e}$)  of the ionised sphere  and the angular
diameter ($\theta_{\rm d}$), temperature  (T$_{\rm d}$), grain opacity
index ($\beta$),  and reference opacity ($\tau_{\rm  1.3\,mm}$) of the
dust  emission.   Since  we  have  seven data  points  with  which  to
constrain  seven  free  parameters,  we  explore  parameter  space  by
constructing   a   40  by   40   point   grid  of   electron   density
(10$^{5.6}$--10$^{7.2}$\,cm$^{-3}$)  and   R$_{e}$  (15--65\,au),  and
fitting  for  the  other  five parameters  (n$_e$,  $\theta_{\rm  d}$,
T$_{\rm d}$, $\beta$,  and $\tau_{\rm 1.3\,mm}$) at each  point in the
n$_e$-R$_e$ grid \citep[see also][]{hunter_2014}.   The fit ranges for
the free  parameters were  T$_e=6000$--$11000$\,K, $\beta=0.5$--$2.5$,
$\theta_{\rm  d}=0.2$--$2.0$\arcsec,  and  T$_{\rm  d}=150$--$250$\,K;
$\tau_{\rm 1.3\,mm}$ was unconstrained.  The range for T$_{\rm d}$ was
chosen   based   on  our   modelling   of   the  CH$_{3}$CN   emission
(Section~\ref{dis:cassis}).  At high density, as  in MM1, gas and dust
temperatures       are       expected        to       be       coupled
\citep[e.g.][]{ceccarelli_1996,kaufman_1998};   since    $\beta$   and
T$_{\rm d}$ are degenerate, we  use the gas temperature measurement to
better constrain the dust model.

\smallskip

The combined  model of a  uniform-density ionised sphere  and graybody
dust    emission   provides    a    good   fit    to   the    observed
centimetre-submillimetre SED of MM1 at wavelengths $\lambda \leq 3$~cm
($\nu\ge$10  GHz),  as  shown   by  the  dashed  blue  curve  in
  Figure~\ref{fig:sed}.    The  best-fit   parameters  of   the  dust
component   are   T$_{\rm  d}=172$\,K,   $\beta=2.1$,   $\tau_{\mathrm
  1.3\,mm}=0.86$, and  $\theta_{\rm d}=0\farcs20\sim700$\,au.  Because
the  primary aim  of our  SED modelling  is to  better understand  the
nature of  the central ionised  source, we consider the  best-fit dust
parameters as indicative, noting  in particular the degeneracy between
dust temperature  and $\beta$ due  to having measurements only  on the
Rayleigh-Jeans portion of the  dust emission spectrum.  Overlaying the
flux  densities  of  \citet{moscadelli_2016}  in  Figure~\ref{fig:sed}
(open  points),  we find  reasonable  consistency  at 2.3  and  1.4~cm
considering the difference  in uv range, but an excess  of emission at
4.8--5.5\,cm compared  to our best-fit dust  and uniform ionised
sphere model.

\smallskip

Given   the    observed   spectral   index   at    these   wavelengths
\citep[$\alpha=0.57\pm0.63$;][]{moscadelli_2016}, we model this excess
as an ionised jet in the form of a truncated power law density profile
($F_\nu \propto \nu^{0.6}$).   Fitting to all of  the datapoints shown
in  Figure~\ref{fig:sed}  (and   listed  in  Table~\ref{tab:sed}),  we
compare three  models: (i) a  three component model including  dust, a
uniform density ionised  sphere (which we interpret  as a hypercompact
(HC) H\,{\sc ii} region, as discussed below), and an ionised jet; (ii)
a  two component  model  including  only dust  and  a uniform  density
ionised sphere;  and (iii) a  two component model including  only dust
and  an  ionised  jet.   For  the  three  component  model,  the  dust
parameters were fixed to those described  above, in light of the large
number of  free parameters.   We first  attempted to  fit for  all six
ionised gas parameters (three describing the ionised sphere, and three
for  the  jet),  but  found  that   the  jet  component  is  not  well
constrained, especially  considering the  dispersion in  the published
flux density measurements at 4.3--5.5\,cm, which is due in part to the
complexity in  the deconvolution of  the bright UC~H\,{\sc  ii} region
(G11.94-0.62)    located     $\sim1'$    to     the    north-northeast
\citep{Wood1989apjs}.  For  this reason, we adopted  a nominal central
density  of  $n_e  =  10^6$\,cm$^{-3}$  and  electron  temperature  of
$T_e$=10$^4$~K and fit only for the  size of the jet.  To assess which
of the  three models --- (i),  (ii), or (iii), above  --- provides the
best description of  the data, we computed the  reduced $\chi^{2}$ for
the  best-fit model  of each  class.  The  three-component model  best
represents the data,  with a reduced $\chi^{2}$ of 1.4,  compared to a
reduced $\chi^{2}$ of 3.0 for the  model with only dust and an ionised
sphere, and a  reduced $\chi^{2}$ of 7.0 for the  model with only dust
and an ionised jet.   The dust and jet model is shown  as a dashed red
curve  in  Figure~\ref{fig:sed},  which illustrates  that  this  model
produces a poor overall fit to the SED, failing to reproduce the $\sim
1.3$\,cm and $\leq 1.1$\,mm  emission.  For the three-component model,
the  best-fit   ionised  gas   parameters  are   electron  temperature
$T_e=9500$\,K, electron density $n_e=5.8 \times 10^{6}$\,cm$^{-3}$ and
radius $R_e$=21\,au ($\sim$ 0\farcs006 at a distance of 3.37\,kpc) for
the  HC~H\,{\sc  ii}  region,  and   a  half-power  radius  of  17\,au
($\sim$0\farcs005   at  3.37\,kpc)   for   the  jet.    As  shown   in
Figure~\ref{fig:sed},  the   three-component  combined   model  (whose
individual components are shown as dotted lines) produces a reasonable
fit to  the SED.  The modelled  source sizes for both  the ionised and
dust  components   are  consistent  with   observational  constraints,
e.g.\ with CM1 being unresolved in  all of our VLA observations and in
those        of        \citet{moscadelli_2016}        (see        also
Section~\ref{sec:cont_results}), and with MM1  being unresolved by the
SMA \citep{cyganowski_2014}.

\smallskip

Our   modelling    results   support   a   picture    in   which   the
centimetre-wavelength  continuum emission  associated with  MM1 arises
from  a  very small  hypercompact  (HC)  H\,{\sc  ii} region  that  is
gravitationally      `trapped'      by     an      accretion      flow
\citep[e.g.][]{keto_2003,keto_2007}, possibly accompanied by a compact
ionised jet.   These results  are consistent  with other  evidence for
ongoing accretion by the central  (proto)star, including the fact that
it drives  an active outflow (Figure  \ref{fig:outflow}).  Indeed, the
momentum outflow  rate estimated by \citet{moscadelli_2016}  from VLBA
observations    of   H$_{2}$O    masers    is   exceptionally    high,
$2\times10^{-2}$    M$_{\odot}$\,yr$^{-1}$\,km\,s$^{-1}$.    A    high
accretion  rate could  also  explain the  moderate  luminosity of  the
region  ($\sim$10$^{4}$\,L$_{\odot}$) compared  to  the enclosed  mass
implied      for     MM1      by      our     kinematic      modelling
(Section~\ref{sec:kinematics}):  in  evolutionary  models  of  massive
(proto)stars,  high accretion  rates  result in  large  radii and  low
effective   temperatures  \citep[e.g.][]{hosokawa_2009,hosokawa_2010}.
Finally,  the early  evolutionary stage  implied by  a gravitationally
trapped H\,{\sc  ii} region and  a swollen, non-ZAMS  configuration is
consistent with the short dynamical timescale of the outflow driven by
MM1, $\lesssim$10,000 years \citep{cyganowski_2011sma}.

\smallskip

The self-consistent scenario  outlined above --- in which  a high rate
of accretion governs  many observable characteristics ---  leads us to
favour   a    HC   H\,{\sc   ii}   region    interpretation   of   the
centimetre-wavelength ($\lambda  = 0.9-3$\,cm) emission from  MM1.  We
note, however, that  HC H\,{\sc ii} regions and ionised  winds or jets
can be difficult to differentiate observationally.  The distinction is
in  part dynamical  --- jets  have  higher velocities  than winds  (as
reflected  in larger  radio  recombination line  widths and/or  proper
motions), which  have higher  velocities than  HC H\,{\sc  ii} regions
\citep[e.g.][]{hoare_2007,hoare_franco_2007}.   The other  distinction
is  the  source  of  ionising  photons: for  HC  H\,{\sc  ii}  regions
(including ionised  accretion flows or discs),  photoionisation by the
central massive (proto)star dominates, while in jets and winds, shocks
may      contribute      significantly     to      the      ionisation
\citep[e.g.][]{shepherd_2004,keto_2007,galvan-madrid_2010}.    Neither
distinguishing  property  can  be   directly  accessed  with  existing
observations, and  both HC  H\,{\sc ii} regions  and ionised  winds or
jets are characterised  by intermediate centimetre-wavelength spectral
indices   (S$_\nu  \propto   \nu^\alpha$,  $-0.1   <  \alpha   <  2$).
\citet{moscadelli_2016}  interpret the  centimetre-wavelength emission
from G11.92--0.61 MM1  as arising from an ionised  wind, based largely
on  their   measured  centimetre-wavelength  spectral   indices.   The
difference  in interpretation  likely  stems from  our more  extensive
wavelength coverage (e.g. into the submillimetre) and the inclusion of
a dust component  in our modelling: a  three-component combined model,
including dust, an HC H\,{\sc ii} region, and an ionised jet, provides
the  best  description  of  the  observed  SED,  as  discussed  above.
\citet{moscadelli_2016}  also  report  a slight  elongation  in  their
highest-resolution (1.4\,cm) image, but  the continuum emission is not
well-resolved (see  also Section~\ref{sec:cont_results}).  Additional,
higher  angular  resolution VLA  observations  (e.g.\  at 0.9\,cm  and
0.7\,cm,  in  the  most-extended  A-configuration)  are  necessary  to
establish whether or  not the ionised component  is spatially extended
on  $\sim$0\farcs1   scales.   Importantly,  a  contribution   to  the
centimetre-wavelength emission from  an ionised wind or jet  --- as in
our  three  component  model (Figure~\ref{fig:sed})  ---  is  entirely
consistent with  our conclusion that  the central  source of MM1  is a
very young massive (proto)star, characterised by ongoing accretion and
a swollen, non-ZAMS configuration.

\subsection{Physical properties of the disc estimated from dust emission}
\label{sec:dust_mass}

Our  modelling   of  MM1's  centimetre-submillimetre   wavelength  SED
confirms  that  the observed  1.3\,mm  flux  density is  dominated  by
thermal dust emission:  in our best-fit model, dust  accounts for 99.5
per cent of the  emission at 1.3\,mm.  We estimate a  gas mass for the
disc from the measured 1.3\,mm  integrated flux density using a simple
model  of  isothermal  dust   emission,  corrected  for  dust  opacity
\citep[][Equation        3]{cyganowski_2011sma}.          As        in
\citet{cyganowski_2014},  these estimates  assume  a gas-to-dust  mass
ratio     of    100     and     a     dust    opacity     $\kappa_{\rm
  1.3\,mm}=1.1$\,cm$^{2}$\,g$^{-1}$  \citep[for grains  with thin  ice
  mantles and  coagulation at 10$^{8}$\,cm$^{-3}$;][]{ossenkopf_1994}.
For these estimates, we adopt the average temperatures of the cool and
warm   components  from   our  pixel-by-pixel   CH$_{3}$CN  modelling,
153$\pm$10\,K and 227$\pm$14\,K  (Section~\ref{dis:cassis}; the quoted
uncertainty is the standard deviation over all fitted pixels), and the
best-fit dust temperature from Section~\ref{sec:ff_sed} of 172\,K.  As
these  temperatures  are   very  similar  to  the   range  adopted  by
\citet{cyganowski_2014} based  on fitting  the CH$_{3}$CN  spectrum at
the millimetre continuum peak, the mass estimates are nearly identical
to those in  \citet{cyganowski_2014}: M$_{\rm gas}=$ 3.3\,M$_{\odot}$,
2.9\,M$_{\odot}$,  and  2.1\,M$_{\odot}$  for  T$_{\rm  dust}=153$\,K,
172\,K,  and  227\,K,   respectively.   We  note  that   even  if  the
centimetre-wavelength  continuum  emission   arose  entirely  from  an
ionised wind \citep[e.g.][]{moscadelli_2016}  with $\alpha=0.6$ out to
millimetre wavelengths,  the contribution to the  1.3\,mm flux density
would be negligible for  our mass estimates ($\sim$1\,mJy contribution
at 1.3\,mm,  corresponding to  a difference in  the estimated  mass of
$\sim0.02$--$0.04$\,\msun, depending on T$_{\rm dust}$).

\smallskip

Calculating corresponding  H$_2$ number  densities for  a cylindrical,
rather  than  a  spherical,   geometry  yields  estimates  of  n$_{\rm
  H_2}\sim$ 2--$3\times$10$^{10}$\,cm$^{-3}$ for a characteristic disc
height of 7\,au  and n$_{\rm H_2}\sim$ 5--$8\times$10$^{9}$\,cm$^{-3}$
for  a  disc  height  of  30\,au (measured  from  the  midplane).   We
emphasise that these estimated number  densities are averages over the
entire  disc,  based  on  an  isothermal  estimate  of  the  gas  mass
associated with the observed dust  emission --- which, as discussed in
Section~\ref{sec:results_extent}, is likely more  compact than the gas
disc.   We  also  note  that  while  the  calculated  dust  opacities,
estimated  as  $\tau_{\rm  dust}=-\mathrm{ln}(1- T_{\rm  b}  /  T_{\rm
  dust})$, are  moderate \citep[0.2--0.3, comparable to  the estimates
  in][]{cyganowski_2014}, these are similarly averages over the entire
source, and do not capture variations in opacity within the disc (e.g.
associated with a dense midplane).

\subsection{G11.92--0.61 MM1 in context}
\label{sec:context}

The  results of  our kinematic  fitting (Section~\ref{sec:kinematics})
suggest that G11.92$-$0.61 MM1 may be the most massive proto-O star to
date with  strong evidence for the  presence of a Keplerian  disc.  Of
candidates  reported in  the  literature, only  AFGL  2591-VLA3 has  a
comparable     central    source     mass:    $\sim$40     M$_{\odot}$
\citep[L$\sim$2$\times$10$^{5}$\,L$_{\odot}$;][]{jimenez-serra_2012,
  sanna_2012}, compared  to $\sim$30-60 M$_{\odot}$ for  MM1.  Studies
of other proto-O stars with evidence for Keplerian discs find enclosed
or central  source masses of $<$30\,M$_{\odot}$.   For NGC6334I(N)-SMA
1b,  perhaps  the  closest  analogue  to  G11.92$-$0.61  based  on  IR
properties,  \citet{hunter_2014}  find  an   enclosed  mass  of  10-30
M$_{\odot}$.  Interestingly, IRAS 16547-4247 and AFGL 4176 --- regions
with     luminosities    6-10$\times$     that    of     G11.92$-$0.61
\citep[$\sim$10$^{4}$\,L$_{\odot}$;][]{cyganowski_2011sma,moscadelli_2016}
--- also have enclosed or  central source masses of $<$30\,M$_{\odot}$
\citep{zapata_2015,johnston_2015}.    Taken   at  face   value,   this
collection  of results  suggests  that enclosed  mass  does not  scale
directly with luminosity for  proto-O stars.  However, the differences
between the estimated luminosities of different sources are comparable
to the uncertainties  in the estimates, particularly  for sources that
do not  have maser parallax  distances (e.g. IRAS 16547-4247  and AFGL
4176).

\smallskip

In contrast, our  estimated (gas) mass for the MM1  disc is similar to
the mass estimates  obtained for other candidate  discs around proto-O
stars using similar  dust properties. In the context  of comparing the
small number of  discs around O-type (proto)stars, it  is worth noting
the  substantial uncertainty  in  mass estimates  associated with  the
(assumed) dust opacity, $\kappa_\nu$.  Our estimates above \citep[like
  those of][]{wang_2012,hunter_2014,zapata_2015}  adopt dust opacities
from  \citet{ossenkopf_1994}   for  grains  with  ice   mantles.   For
\citet{draine_2003}  interstellar  grains,  $\kappa_{\rm 1.2  mm}$  is
lower  by a  factor  of  $\gtrsim$5 (for  R$_{V}=5.5$,  as adopted  by
\citealt{johnston_2015}), yielding  mass estimates that are  larger by
the same factor.  This difference  in assumed dust properties accounts
for the  much larger  mass reported  by \citet{johnston_2015}  for the
AFGL 4176 disc ($\sim$12 M$_{\odot}$), compared to the values of a few
solar masses  ($\sim$2-6 M$_{\odot}$) characteristic of  MM1 and other
massive                         disc                        candidates
\citep[e.g.][]{wang_2012,hunter_2014,zapata_2015}. 

\smallskip

The  relatively  large  disc-to-star   mass  ratio  derived  from  our
observations  of  G11.92$-$0.61  MM1 ($\gtrsim$0.035)  indicates  that
self-gravity  may  play   a  role  in  the  evolution   of  the  disc.
Self-gravitating discs are efficient transporters of angular momentum,
and hence  provide a suitable  means of assembling  relatively massive
stars  on short  timescales.   In  a companion  paper  (Forgan et  al,
submitted) we compute simple  semi-analytic models of self-gravitating
discs  \citep{forgan_2011a,forgan_2013}, both  for MM1  and for  other
massive  Keplerian disc  candidates recently  observed.  We  find that
these simple models provide reasonably  good estimates of the observed
disc mass, given  the observational constraints on the  disc inner and
outer radii, stellar mass and  accretion rate.  Most intriguingly, the
models  predict  that  the  disc around  MM1  should  be  sufficiently
unstable  to fragment  into low  mass protostars.   These objects  are
beyond the  resolution of our  observations at  this time, but  may be
detectable with e.g.\ ALMA.

\section{Conclusions}
\label{sec:conclusions}

\begin{figure}   % Fig 9
  \includegraphics[width=\columnwidth]{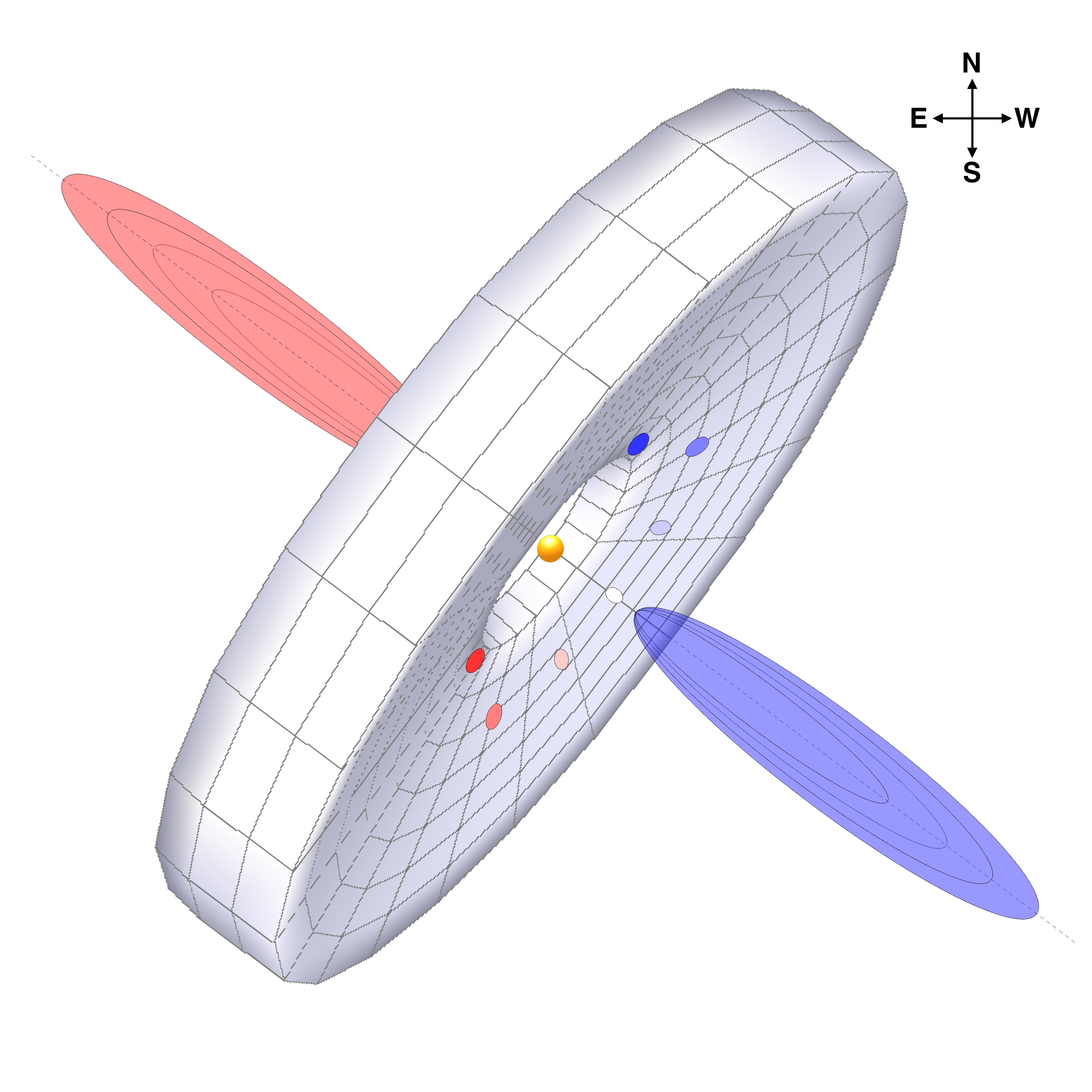}
  \caption{A proposed morphology for the  immediate vicinity of MM1 as
    viewed from  Earth.  The  maximum velocity gradient  obtained from
    the line channel analysis  (red-blue points) lies perpendicular to
    the axis of the  molecular outflow (Fig.  \ref{fig:outflow}).  The
    closed    loop     structure    of    the     channel    centroids
    (Fig. \ref{fig:all_centroids}) for many  of the emission lines can
    be explained by  a flared, optically thick  Keplerian disc, viewed
    at an intermediate inclination.  Such  a disc would also exhibit a
    hot    spot     in    temperature    towards     the    South-West
    (Fig. \ref{fig:ch3cn_cassis}).}
    \label{fig:morph}
\end{figure}

In  this  paper,  we   have  used  sub-arcsecond  ($\lesssim$1550\,au)
resolution  SMA 1.3  mm and  VLA 3.0  and 0.9  cm observations  of the
high-mass (proto)star  G11.92--0.61 MM1 to characterise  the nature of
the central source and examine  the kinematics and physical conditions
of the gas surrounding it.  Our main findings are:

\begin{itemize}
  
  \item The  compact molecular line  emission associated with  the MM1
    millimetre continuum  source exhibits a velocity  gradient that is
    approximately perpendicular to the high-velocity bipolar molecular
    outflow driven by  MM1.  The velocity gradient of  the compact gas
    is  remarkably   consistent,  being  seen  in   lines  of  varying
    excitation energy and in  10 different chemical species, including
    CH$_{3}$OH, CH$_{3}$OCHO, OCS, HNCO, CH$_{3}$CH$_{2}$CN, H$_{2}$CO
    and DCN.
  
  \smallskip
  
  \item From the  8 GHz of bandwidth covered by  our SMA observations,
    we identify 15 potentially `disc tracing' lines.  For all of these
    lines,  position-velocity  diagrams   exhibit  the  characteristic
    pattern expected for a Keplerian  disc: the highest velocities are
    seen nearest the  central source, and the  most spatially extended
    emission at lower velocities.
  
  \smallskip
  
  \item  Our kinematic  modelling of  the MM1  disc yields  a best-fit
    enclosed   mass   of    $60^{+21}_{-27}$\,M$_{\odot}$   and   disc
    inclination of $35^{+20}_{-6}\degr$ (to  the line of sight), using
    all `disc-tracing' lines other than CH$_{3}$OH (nine transitions).
    The  six  `disc-tracing'  transitions   of  CH$_{3}$OH  exhibit  a
    somewhat different morphology than the  rest of the lines in plots
    of emission centroid position as  a function of velocity, and were
    fit separately.  The best  fitting disc  model for  the CH$_{3}$OH
    transitions    yields    a    disc    enclosing    a    mass    of
    $34^{+28}_{-12}$\,M$_{\odot}$     at     an     inclination     of
    $52^{+11}_{-14}\degr$.   Together these  results  imply a  central
    source mass  of $\sim$30--60  M$_{\odot}$, making  MM1 potentially
    the most massive proto-O star with strong evidence for a Keplerian
    disc.

   \smallskip
    
   \item  For many  observed molecular  lines, the  extent of  the gas
     emission is larger than that  of the dust continuum emission from
     the disc.  From  our kinematic modelling of the  gas emission, we
     find a disc outer radius of 1200\,au.
    
    \smallskip

    \item  Two   temperature  components  are  required   to  fit  the
      CH$_{3}$CN  J=12-11 emission  from  the MM1  disc  using an  LTE
      radiative transfer model.  The cooler ($153\pm10$\,K) and warmer
      ($227\pm14$\,K)   components   may  correspond   to   CH$_{3}$CN
      reservoirs  in different  disc layers,  as in  recent models  of
      discs around low-mass protostars \citep{walsh_2015}.

   \smallskip

    \item Our modelling of the centimetre-submillimetre wavelength SED
      of  MM1  confirms that  the  observed  1.3\,mm flux  density  is
      dominated by dust.   Applying a simple model  of isothermal dust
      emission, we  estimate a disc gas  mass of 2.1--3.3\,M$_{\odot}$
      (for T$_{\rm dust}=227$--153\,K).  This mass estimate is similar
      to those for most other candidate Keplerian discs around proto-O
      stars  that assume  similar dust  properties.  MM1's  relatively
      high disc-to-star mass ratio ($\gtrsim$ 0.035) suggests that the
      disc may  be self-gravitating, which  we explore in  a companion
      modelling paper (Forgan et al., submitted).
    
  \smallskip
  
  \item From our SED modelling, we find that the centimetre-wavelength
    flux density of MM1 is  dominated by free-free emission.  We model
    the free-free and dust emission  simultaneously, and find that the
    ionised gas  is best-fit by a  model of a uniform  density ionised
    sphere  with   electron  temperature  9500\,K,   electron  density
    $5.8\times10^{6}$\,cm$^{-3}$ and radius  21\,au.  These properties
    are consistent with a very small, gravitationally trapped HC H{\sc
      ii} region, possibly accompanied by a compact ionised jet.
     
\end{itemize}

In combination, our results suggest  that G11.92--0.61 MM1 is likely a
young proto-O  star, in a swollen,  non-ZAMS configuration, surrounded
by  a Keplerian  disc  with  a morphology  similar  to  that shown  in
Figure~\ref{fig:morph}.   Our  observations  and modelling  support  a
self-consistent  picture in  which accretion  is ongoing,  and a  high
accretion  rate  governs  many observable  properties,  including  the
presence  of a  gravitationally  trapped  HC H{\sc  ii}  region and  a
moderate  luminosity  ($\sim$  10$^{4}$  L$_{\odot}$)  for  a  massive
($\sim$30-60M$_{\odot}$)   central   star.   Future   higher   angular
resolution  observations will  be  required to  spatially resolve  the
protostar-disc system.  Such observations will be essential to compare
the role  and physics  of disc  accretion in  high- and  low-mass star
formation, and  to develop  a clearer  understanding of  the accretion
processes at work in massive young stellar objects.  
 
\section*{Acknowledgements}

We  would like  to  thank  Melvin Hoare,  Simon  Purser and  Katharine
Johnston  for   helpful  discussions  regarding  the   nature  of  the
centimetre-wavelength  emission,   and  Catherine  Walsh   for  kindly
providing further  information on the  behaviour of CH$_{3}$CN  in her
chemical  models.   We  also  thank Kenny  Wood,  Claire  Davies,  and
Christine Koepferl for their input at  an early stage of this project,
and the referee for a constructive report.

\smallskip

JDI gratefully  acknowledges support  from the DISCSIM  project, grant
agreement  341137,  funded  by  the European  Research  Council  under
ERC-2013-ADG.  CJC acknowledges support  from STFC grant ST/M001296/1.
PN,  CJC and  JDI gratefully  acknowledge support  in the  form of  an
Undergraduate Research Bursary from the Royal Astronomical Society. DF
acknowledges support from the  ECOGAL project, grant agreement 291227,
funded  by the  European  Research Council  under ERC-2011-ADG.   This
research has made use of NASA's Astrophysics Data System Bibliographic
Services;  Astropy,  a  community-developed core  Python  package  for
Astronomy \citep{astropy_2013}; APLpy, an open-source plotting package
for  Python hosted  at \url{http://aplpy.github.com},  and the  CASSIS
software and VADMC databases (\url{http://www.vamdc.eu/}).  CASSIS has
been  developed  by  IRAP-UPS/CNRS  (\url{http://cassis.irap.omp.eu}).
The Submillimeter  Array is  a joint  project between  the Smithsonian
Astrophysical  Observatory  and  the   Academia  Sinica  Institute  of
Astronomy  and   Astrophysics,  and  is  funded   by  the  Smithsonian
Institution  and the  Academia Sinica.   The National  Radio Astronomy
Observatory is a facility of  the National Science Foundation operated
under agreement by the Associated Universities, Inc.



\bibliographystyle{mnras} 
\bibliography{mm1}

\begin{thebibliography}{}
\makeatletter
\relax
\def\mn@urlcharsother{\let\do\@makeother \do\$\do\&\do\#\do\^\do\_\do\%\do\~}
\def\mn@doi{\begingroup\mn@urlcharsother \@ifnextchar [ {\mn@doi@}
  {\mn@doi@[]}}
\def\mn@doi@[#1]#2{\def\@tempa{#1}\ifx\@tempa\@empty \href
  {http://dx.doi.org/#2} {doi:#2}\else \href {http://dx.doi.org/#2} {#1}\fi
  \endgroup}
\def\mn@eprint#1#2{\mn@eprint@#1:#2::\@nil}
\def\mn@eprint@arXiv#1{\href {http://arxiv.org/abs/#1} {{\tt arXiv:#1}}}
\def\mn@eprint@dblp#1{\href {http://dblp.uni-trier.de/rec/bibtex/#1.xml}
  {dblp:#1}}
\def\mn@eprint@#1:#2:#3:#4\@nil{\def\@tempa {#1}\def\@tempb {#2}\def\@tempc
  {#3}\ifx \@tempc \@empty \let \@tempc \@tempb \let \@tempb \@tempa \fi \ifx
  \@tempb \@empty \def\@tempb {arXiv}\fi \@ifundefined
  {mn@eprint@\@tempb}{\@tempb:\@tempc}{\expandafter \expandafter \csname
  mn@eprint@\@tempb\endcsname \expandafter{\@tempc}}}

\bibitem[\protect\citeauthoryear{{Alexander}, {Clarke}  \&
  {Pringle}}{{Alexander} et~al.}{2006}]{alexander_2006}
{Alexander} R.~D.,  {Clarke} C.~J.,   {Pringle} J.~E.,  2006, \mn@doi [\mnras]
  {10.1111/j.1365-2966.2006.10294.x}, \href
  {http://adsabs.harvard.edu/abs/2006MNRAS.369..229A} {369, 229}

\bibitem[\protect\citeauthoryear{{Araya}, {Hofner}, {Kurtz}, {Bronfman}  \&
  {DeDeo}}{{Araya} et~al.}{2005}]{araya_2005}
{Araya} E.,  {Hofner} P.,  {Kurtz} S.,  {Bronfman} L.,   {DeDeo} S.,  2005,
  \mn@doi [\apjs] {10.1086/427187}, \href
  {http://adsabs.harvard.edu/abs/2005ApJS..157..279A} {157, 279}

\bibitem[\protect\citeauthoryear{{Astropy Collaboration} et~al.,}{{Astropy
  Collaboration} et~al.}{2013}]{astropy_2013}
{Astropy Collaboration} et~al., 2013, \mn@doi [\aap]
  {10.1051/0004-6361/201322068}, \href
  {http://adsabs.harvard.edu/abs/2013A%26A...558A..33A} {558, A33}

\bibitem[\protect\citeauthoryear{{Beltr{\'a}n} \& {de Wit}}{{Beltr{\'a}n} \&
  {de Wit}}{2016}]{beltran_2016}
{Beltr{\'a}n} M.~T.,  {de Wit} W.~J.,  2016, \mn@doi [\aapr]
  {10.1007/s00159-015-0089-z}, \href
  {http://adsabs.harvard.edu/abs/2016A%26ARv..24....6B} {24, 6}

\bibitem[\protect\citeauthoryear{{Beltr{\'a}n}, {Cesaroni}, {Neri}  \&
  {Codella}}{{Beltr{\'a}n} et~al.}{2011}]{beltran_2011}
{Beltr{\'a}n} M.~T.,  {Cesaroni} R.,  {Neri} R.,   {Codella} C.,  2011, \mn@doi
  [\aap] {10.1051/0004-6361/201015049}, \href
  {http://adsabs.harvard.edu/abs/2011A%26A...525A.151B} {525, A151}

\bibitem[\protect\citeauthoryear{{Beltr{\'a}n} et~al.,}{{Beltr{\'a}n}
  et~al.}{2014}]{beltran_2014}
{Beltr{\'a}n} M.~T.,  et~al., 2014, \mn@doi [\aap]
  {10.1051/0004-6361/201424031}, \href
  {http://adsabs.harvard.edu/abs/2014A%26A...571A..52B} {571, A52}

\bibitem[\protect\citeauthoryear{{Benedettini} et~al.,}{{Benedettini}
  et~al.}{2013}]{benedettini_2013}
{Benedettini} M.,  et~al., 2013, \mn@doi [\mnras] {10.1093/mnras/stt1559},
  \href {http://adsabs.harvard.edu/abs/2013MNRAS.436..179B} {436, 179}

\bibitem[\protect\citeauthoryear{{Beuther} \& {Walsh}}{{Beuther} \&
  {Walsh}}{2008}]{beuther_2008}
{Beuther} H.,  {Walsh} A.~J.,  2008, \mn@doi [\apjl] {10.1086/527434}, \href
  {http://adsabs.harvard.edu/abs/2008ApJ...673L..55B} {673, L55}

\bibitem[\protect\citeauthoryear{{Bik} \& {Thi}}{{Bik} \&
  {Thi}}{2004}]{bik_2004}
{Bik} A.,  {Thi} W.~F.,  2004, \mn@doi [\aap] {10.1051/0004-6361:200400088},
  \href {http://adsabs.harvard.edu/abs/2004A%26A...427L..13B} {427, L13}

\bibitem[\protect\citeauthoryear{{Boley} et~al.,}{{Boley}
  et~al.}{2013}]{boley_2013}
{Boley} P.~A.,  et~al., 2013, \mn@doi [\aap] {10.1051/0004-6361/201321539},
  \href {http://adsabs.harvard.edu/abs/2013A%26A...558A..24B} {558, A24}

\bibitem[\protect\citeauthoryear{{Breen} \& {Ellingsen}}{{Breen} \&
  {Ellingsen}}{2011}]{breen_2011}
{Breen} S.~L.,  {Ellingsen} S.~P.,  2011, \mn@doi [\mnras]
  {10.1111/j.1365-2966.2011.19020.x}, \href
  {http://adsabs.harvard.edu/abs/2011MNRAS.416..178B} {416, 178}

\bibitem[\protect\citeauthoryear{{Brogan}, {Chandler}, {Hunter}, {Shirley}  \&
  {Sarma}}{{Brogan} et~al.}{2007}]{brogan_2007}
{Brogan} C.~L.,  {Chandler} C.~J.,  {Hunter} T.~R.,  {Shirley} Y.~L.,   {Sarma}
  A.~P.,  2007, \mn@doi [\apjl] {10.1086/518390}, \href
  {http://adsabs.harvard.edu/abs/2007ApJ...660L.133B} {660, L133}

\bibitem[\protect\citeauthoryear{{Ceccarelli}, {Hollenbach}  \&
  {Tielens}}{{Ceccarelli} et~al.}{1996}]{ceccarelli_1996}
{Ceccarelli} C.,  {Hollenbach} D.~J.,   {Tielens} A.~G.~G.~M.,  1996, \mn@doi
  [\apj] {10.1086/177978}, \href
  {http://adsabs.harvard.edu/abs/1996ApJ...471..400C} {471, 400}

\bibitem[\protect\citeauthoryear{{Cesaroni}}{{Cesaroni}}{2005}]{cesaroni_2005_apss}
{Cesaroni} R.,  2005, \mn@doi [\apss] {10.1007/s10509-005-3651-8}, \href
  {http://adsabs.harvard.edu/abs/2005Ap%26SS.295....5C} {295, 5}

\bibitem[\protect\citeauthoryear{{Cesaroni}, {Galli}, {Lodato}, {Walmsley}  \&
  {Zhang}}{{Cesaroni} et~al.}{2006}]{cesaroni_2006}
{Cesaroni} R.,  {Galli} D.,  {Lodato} G.,  {Walmsley} M.,   {Zhang} Q.,  2006,
  \mn@doi [\nat] {10.1038/nature05344}, \href
  {http://adsabs.harvard.edu/abs/2006Natur.444..703C} {444, 703}

\bibitem[\protect\citeauthoryear{{Cesaroni}, {Galli}, {Lodato}, {Walmsley}  \&
  {Zhang}}{{Cesaroni} et~al.}{2007}]{cesaroni_2007}
{Cesaroni} R.,  {Galli} D.,  {Lodato} G.,  {Walmsley} C.~M.,   {Zhang} Q.,
  2007, Protostars and Planets V, \href
  {http://adsabs.harvard.edu/abs/2007prpl.conf..197C} {pp 197--212}

\bibitem[\protect\citeauthoryear{{Cesaroni}, {Beltr{\'a}n}, {Zhang}, {Beuther}
  \& {Fallscheer}}{{Cesaroni} et~al.}{2011}]{cesaroni_2011}
{Cesaroni} R.,  {Beltr{\'a}n} M.~T.,  {Zhang} Q.,  {Beuther} H.,   {Fallscheer}
  C.,  2011, \mn@doi [\aap] {10.1051/0004-6361/201117206}, \href
  {http://adsabs.harvard.edu/abs/2011A%26A...533A..73C} {533, A73}

\bibitem[\protect\citeauthoryear{{Cesaroni}, {Galli}, {Neri}  \&
  {Walmsley}}{{Cesaroni} et~al.}{2014}]{cesaroni_2014}
{Cesaroni} R.,  {Galli} D.,  {Neri} R.,   {Walmsley} C.~M.,  2014, \mn@doi
  [\aap] {10.1051/0004-6361/201323065}, \href
  {http://adsabs.harvard.edu/abs/2014A%26A...566A..73C} {566, A73}

\bibitem[\protect\citeauthoryear{{Comito}, {Schilke}, {Endesfelder},
  {Jim{\'e}nez-Serra}  \& {Mart{\'{\i}}n-Pintado}}{{Comito}
  et~al.}{2007}]{comito_2007}
{Comito} C.,  {Schilke} P.,  {Endesfelder} U.,  {Jim{\'e}nez-Serra} I.,
  {Mart{\'{\i}}n-Pintado} J.,  2007, \mn@doi [\aap]
  {10.1051/0004-6361:20077408}, \href
  {http://adsabs.harvard.edu/abs/2007A%26A...469..207C} {469, 207}

\bibitem[\protect\citeauthoryear{{Cyganowski} et~al.,}{{Cyganowski}
  et~al.}{2008}]{cyganowski_2008}
{Cyganowski} C.~J.,  et~al., 2008, \mn@doi [\aj]
  {10.1088/0004-6256/136/6/2391}, \href
  {http://adsabs.harvard.edu/abs/2008AJ....136.2391C} {136, 2391}

\bibitem[\protect\citeauthoryear{{Cyganowski}, {Brogan}, {Hunter}  \&
  {Churchwell}}{{Cyganowski} et~al.}{2009}]{cyganowski_2009}
{Cyganowski} C.~J.,  {Brogan} C.~L.,  {Hunter} T.~R.,   {Churchwell} E.,  2009,
  \mn@doi [\apj] {10.1088/0004-637X/702/2/1615}, \href
  {http://adsabs.harvard.edu/abs/2009ApJ...702.1615C} {702, 1615}

\bibitem[\protect\citeauthoryear{{Cyganowski}, {Brogan}, {Hunter}, {Churchwell}
   \& {Zhang}}{{Cyganowski} et~al.}{2011a}]{cyganowski_2011sma}
{Cyganowski} C.~J.,  {Brogan} C.~L.,  {Hunter} T.~R.,  {Churchwell} E.,
  {Zhang} Q.,  2011a, \mn@doi [\apj] {10.1088/0004-637X/729/2/124}, \href
  {http://adsabs.harvard.edu/abs/2011ApJ...729..124C} {729, 124}

\bibitem[\protect\citeauthoryear{{Cyganowski}, {Brogan}, {Hunter}  \&
  {Churchwell}}{{Cyganowski} et~al.}{2011b}]{cyganowski_2011vla}
{Cyganowski} C.~J.,  {Brogan} C.~L.,  {Hunter} T.~R.,   {Churchwell} E.,
  2011b, \mn@doi [\apj] {10.1088/0004-637X/743/1/56}, \href
  {http://adsabs.harvard.edu/abs/2011ApJ...743...56C} {743, 56}

\bibitem[\protect\citeauthoryear{{Cyganowski} et~al.,}{{Cyganowski}
  et~al.}{2014}]{cyganowski_2014}
{Cyganowski} C.~J.,  et~al., 2014, \mn@doi [\apjl]
  {10.1088/2041-8205/796/1/L2}, \href
  {http://adsabs.harvard.edu/abs/2014ApJ...796L...2C} {796, L2}

\bibitem[\protect\citeauthoryear{{Davies}, {Lumsden}, {Hoare}, {Oudmaijer}  \&
  {de Wit}}{{Davies} et~al.}{2010}]{davies_2010}
{Davies} B.,  {Lumsden} S.~L.,  {Hoare} M.~G.,  {Oudmaijer} R.~D.,   {de Wit}
  W.-J.,  2010, \mn@doi [\mnras] {10.1111/j.1365-2966.2009.16077.x}, \href
  {http://adsabs.harvard.edu/abs/2010MNRAS.402.1504D} {402, 1504}

\bibitem[\protect\citeauthoryear{{Davies}, {Hoare}, {Lumsden}, {Hosokawa},
  {Oudmaijer}, {Urquhart}, {Mottram}  \& {Stead}}{{Davies}
  et~al.}{2011}]{davies_2011}
{Davies} B.,  {Hoare} M.~G.,  {Lumsden} S.~L.,  {Hosokawa} T.,  {Oudmaijer}
  R.~D.,  {Urquhart} J.~S.,  {Mottram} J.~C.,   {Stead} J.,  2011, \mn@doi
  [\mnras] {10.1111/j.1365-2966.2011.19095.x}, \href
  {http://adsabs.harvard.edu/abs/2011MNRAS.416..972D} {416, 972}

\bibitem[\protect\citeauthoryear{{Draine}}{{Draine}}{2003}]{draine_2003}
{Draine} B.~T.,  2003, \mn@doi [\araa]
  {10.1146/annurev.astro.41.011802.094840}, \href
  {http://adsabs.harvard.edu/abs/2003ARA%26A..41..241D} {41, 241}

\bibitem[\protect\citeauthoryear{Forgan \& Rice}{Forgan \&
  Rice}{2011}]{forgan_2011a}
Forgan D.,  Rice K.,  2011, \mn@doi [MNRAS] {10.1111/j.1365-2966.2011.19380.x},
  417, 1928

\bibitem[\protect\citeauthoryear{Forgan \& Rice}{Forgan \&
  Rice}{2013}]{forgan_2013}
Forgan D.,  Rice K.,  2013, \mn@doi [MNRAS] {10.1093/mnras/stt032}, 430, 2082

\bibitem[\protect\citeauthoryear{{Galv{\'a}n-Madrid}, {Zhang}, {Keto}, {Ho},
  {Zapata}, {Rodr{\'{\i}}guez}, {Pineda}  \&
  {V{\'a}zquez-Semadeni}}{{Galv{\'a}n-Madrid}
  et~al.}{2010}]{galvan-madrid_2010}
{Galv{\'a}n-Madrid} R.,  {Zhang} Q.,  {Keto} E.,  {Ho} P.~T.~P.,  {Zapata}
  L.~A.,  {Rodr{\'{\i}}guez} L.~F.,  {Pineda} J.~E.,   {V{\'a}zquez-Semadeni}
  E.,  2010, \mn@doi [\apj] {10.1088/0004-637X/725/1/17}, \href
  {http://adsabs.harvard.edu/abs/2010ApJ...725...17G} {725, 17}

\bibitem[\protect\citeauthoryear{{Gordon}}{{Gordon}}{1995}]{gordon_1995}
{Gordon} M.~A.,  1995, \aap, \href
  {http://adsabs.harvard.edu/abs/1995A%26A...301..853G} {301, 853}

\bibitem[\protect\citeauthoryear{{Guilloteau}, {Dutrey}, {Pi{\'e}tu}  \&
  {Boehler}}{{Guilloteau} et~al.}{2011}]{guilloteau_2011}
{Guilloteau} S.,  {Dutrey} A.,  {Pi{\'e}tu} V.,   {Boehler} Y.,  2011, \mn@doi
  [\aap] {10.1051/0004-6361/201015209}, \href
  {http://adsabs.harvard.edu/abs/2011A%26A...529A.105G} {529, A105}

\bibitem[\protect\citeauthoryear{{Harries}, {Haworth}  \& {Acreman}}{{Harries}
  et~al.}{2014}]{harries_2014}
{Harries} T.~J.,  {Haworth} T.~J.,   {Acreman} D.~M.,  2014, \mn@doi
  [Astrophysics and Space Science Proceedings] {10.1007/978-3-319-03041-8_77},
  \href {http://adsabs.harvard.edu/abs/2014ASSP...36..395H} {36, 395}

\bibitem[\protect\citeauthoryear{{Hern{\'a}ndez-Hern{\'a}ndez}, {Zapata},
  {Kurtz}  \& {Garay}}{{Hern{\'a}ndez-Hern{\'a}ndez}
  et~al.}{2014}]{hernandez-hernandez_2014}
{Hern{\'a}ndez-Hern{\'a}ndez} V.,  {Zapata} L.,  {Kurtz} S.,   {Garay} G.,
  2014, \mn@doi [\apj] {10.1088/0004-637X/786/1/38}, \href
  {http://adsabs.harvard.edu/abs/2014ApJ...786...38H} {786, 38}

\bibitem[\protect\citeauthoryear{{Hoare} \& {Franco}}{{Hoare} \&
  {Franco}}{2007}]{hoare_franco_2007}
{Hoare} M.~G.,  {Franco} J.,  2007, \mn@doi [Astrophysics and Space Science
  Proceedings] {10.1007/978-1-4020-5425-9_4}, \href
  {http://adsabs.harvard.edu/abs/2007ASSP....1...61H} {1, 61}

\bibitem[\protect\citeauthoryear{{Hoare}, {Kurtz}, {Lizano}, {Keto}  \&
  {Hofner}}{{Hoare} et~al.}{2007}]{hoare_2007}
{Hoare} M.~G.,  {Kurtz} S.~E.,  {Lizano} S.,  {Keto} E.,   {Hofner} P.,  2007,
  Protostars and Planets V, \href
  {http://adsabs.harvard.edu/abs/2007prpl.conf..181H} {pp 181--196}

\bibitem[\protect\citeauthoryear{{Hofner} \& {Churchwell}}{{Hofner} \&
  {Churchwell}}{1996}]{hofner_1996}
{Hofner} P.,  {Churchwell} E.,  1996, \aaps, \href
  {http://cdsads.u-strasbg.fr/abs/1996A%26AS..120..283H} {120, 283}

\bibitem[\protect\citeauthoryear{{Hosokawa} \& {Omukai}}{{Hosokawa} \&
  {Omukai}}{2009}]{hosokawa_2009}
{Hosokawa} T.,  {Omukai} K.,  2009, \mn@doi [\apj]
  {10.1088/0004-637X/691/1/823}, \href
  {http://adsabs.harvard.edu/abs/2009ApJ...691..823H} {691, 823}

\bibitem[\protect\citeauthoryear{{Hosokawa}, {Yorke}  \& {Omukai}}{{Hosokawa}
  et~al.}{2010}]{hosokawa_2010}
{Hosokawa} T.,  {Yorke} H.~W.,   {Omukai} K.,  2010, \mn@doi [\apj]
  {10.1088/0004-637X/721/1/478}, \href
  {http://adsabs.harvard.edu/abs/2010ApJ...721..478H} {721, 478}

\bibitem[\protect\citeauthoryear{{Hughes}, {Wilner}, {Andrews}, {Qi}  \&
  {Hogerheijde}}{{Hughes} et~al.}{2011}]{hughes_2011}
{Hughes} A.~M.,  {Wilner} D.~J.,  {Andrews} S.~M.,  {Qi} C.,   {Hogerheijde}
  M.~R.,  2011, \mn@doi [\apj] {10.1088/0004-637X/727/2/85}, \href
  {http://adsabs.harvard.edu/abs/2011ApJ...727...85H} {727, 85}

\bibitem[\protect\citeauthoryear{{Hunter}, {Brogan}, {Cyganowski}  \&
  {Young}}{{Hunter} et~al.}{2014}]{hunter_2014}
{Hunter} T.~R.,  {Brogan} C.~L.,  {Cyganowski} C.~J.,   {Young} K.~H.,  2014,
  \mn@doi [\apj] {10.1088/0004-637X/788/2/187}, \href
  {http://adsabs.harvard.edu/abs/2014ApJ...788..187H} {788, 187}

\bibitem[\protect\citeauthoryear{{Ilee} et~al.,}{{Ilee}
  et~al.}{2013}]{ilee_2013}
{Ilee} J.~D.,  et~al., 2013, \mn@doi [\mnras] {10.1093/mnras/sts537}, \href
  {http://adsabs.harvard.edu/abs/2013MNRAS.429.2960I} {429, 2960}

\bibitem[\protect\citeauthoryear{{Ilee}, {Fairlamb}, {Oudmaijer},
  {Mendigut{\'{\i}}a}, {van den Ancker}, {Kraus}  \& {Wheelwright}}{{Ilee}
  et~al.}{2014}]{ilee_2014}
{Ilee} J.~D.,  {Fairlamb} J.,  {Oudmaijer} R.~D.,  {Mendigut{\'{\i}}a} I.,
  {van den Ancker} M.~E.,  {Kraus} S.,   {Wheelwright} H.~E.,  2014, \mn@doi
  [\mnras] {10.1093/mnras/stu1942}, \href
  {http://adsabs.harvard.edu/abs/2014MNRAS.445.3723I} {445, 3723}

\bibitem[\protect\citeauthoryear{{Jim{\'e}nez-Serra}, {Zhang}, {Viti},
  {Mart{\'{\i}}n-Pintado}  \& {de Wit}}{{Jim{\'e}nez-Serra}
  et~al.}{2012}]{jimenez-serra_2012}
{Jim{\'e}nez-Serra} I.,  {Zhang} Q.,  {Viti} S.,  {Mart{\'{\i}}n-Pintado} J.,
  {de Wit} W.-J.,  2012, \mn@doi [\apj] {10.1088/0004-637X/753/1/34}, \href
  {http://adsabs.harvard.edu/abs/2012ApJ...753...34J} {753, 34}

\bibitem[\protect\citeauthoryear{{Johnston}, {Beuther}, {Linz}, {Boley},
  {Robitaille}, {Keto}, {Wood}  \& {van Boekel}}{{Johnston}
  et~al.}{2014}]{johnston_2014}
{Johnston} K.~G.,  {Beuther} H.,  {Linz} H.,  {Boley} P.,  {Robitaille} T.~P.,
  {Keto} E.,  {Wood} K.,   {van Boekel} R.,  2014, \mn@doi [Astrophysics and
  Space Science Proceedings] {10.1007/978-3-319-03041-8_80}, \href
  {http://adsabs.harvard.edu/abs/2014ASSP...36..413J} {36, 413}

\bibitem[\protect\citeauthoryear{{Johnston} et~al.,}{{Johnston}
  et~al.}{2015}]{johnston_2015}
{Johnston} K.~G.,  et~al., 2015, \mn@doi [\apjl] {10.1088/2041-8205/813/1/L19},
  \href {http://adsabs.harvard.edu/abs/2015ApJ...813L..19J} {813, L19}

\bibitem[\protect\citeauthoryear{{Kaufman}, {Hollenbach}  \&
  {Tielens}}{{Kaufman} et~al.}{1998}]{kaufman_1998}
{Kaufman} M.~J.,  {Hollenbach} D.~J.,   {Tielens} A.~G.~G.~M.,  1998, \mn@doi
  [\apj] {10.1086/305444}, \href
  {http://adsabs.harvard.edu/abs/1998ApJ...497..276K} {497, 276}

\bibitem[\protect\citeauthoryear{{Keto}}{{Keto}}{2003}]{keto_2003}
{Keto} E.,  2003, \mn@doi [\apj] {10.1086/379545}, \href
  {http://adsabs.harvard.edu/abs/2003ApJ...599.1196K} {599, 1196}

\bibitem[\protect\citeauthoryear{{Keto}}{{Keto}}{2007}]{keto_2007}
{Keto} E.,  2007, \mn@doi [\apj] {10.1086/520320}, \href
  {http://adsabs.harvard.edu/abs/2007ApJ...666..976K} {666, 976}

\bibitem[\protect\citeauthoryear{{Klassen}, {Pudritz}, {Kuiper}, {Peters}  \&
  {Banerjee}}{{Klassen} et~al.}{2016}]{klassen_2016}
{Klassen} M.,  {Pudritz} R.,  {Kuiper} R.,  {Peters} T.,   {Banerjee} R.,
  2016, preprint, \href {http://adsabs.harvard.edu/abs/2016arXiv160307345K} {}
  (\mn@eprint {arXiv} {1603.07345})

\bibitem[\protect\citeauthoryear{{Kraus} et~al.,}{{Kraus}
  et~al.}{2010}]{kraus_2010}
{Kraus} S.,  et~al., 2010, \mn@doi [\nat] {10.1038/nature09174}, \href
  {http://adsabs.harvard.edu/abs/2010Natur.466..339K} {466, 339}

\bibitem[\protect\citeauthoryear{{Krumholz}, {Klein}, {McKee}, {Offner}  \&
  {Cunningham}}{{Krumholz} et~al.}{2009}]{krumholz_2009}
{Krumholz} M.~R.,  {Klein} R.~I.,  {McKee} C.~F.,  {Offner} S.~S.~R.,
  {Cunningham} A.~J.,  2009, \mn@doi [Science] {10.1126/science.1165857}, \href
  {http://adsabs.harvard.edu/abs/2009Sci...323..754K} {323, 754}

\bibitem[\protect\citeauthoryear{{Kuiper}, {Klahr}, {Beuther}  \&
  {Henning}}{{Kuiper} et~al.}{2010}]{kuiper_2010}
{Kuiper} R.,  {Klahr} H.,  {Beuther} H.,   {Henning} T.,  2010, \mn@doi [\apj]
  {10.1088/0004-637X/722/2/1556}, \href
  {http://adsabs.harvard.edu/abs/2010ApJ...722.1556K} {722, 1556}

\bibitem[\protect\citeauthoryear{{Kuiper}, {Klahr}, {Beuther}  \&
  {Henning}}{{Kuiper} et~al.}{2011}]{kuiper_2011}
{Kuiper} R.,  {Klahr} H.,  {Beuther} H.,   {Henning} T.,  2011, \mn@doi [\apj]
  {10.1088/0004-637X/732/1/20}, \href
  {http://adsabs.harvard.edu/abs/2011ApJ...732...20K} {732, 20}

\bibitem[\protect\citeauthoryear{{Lee}, {Takami}, {Duan}, {Karr}, {Su}, {Liu},
  {Froebrich}  \& {Yeh}}{{Lee} et~al.}{2012}]{lee_2012}
{Lee} H.-T.,  {Takami} M.,  {Duan} H.-Y.,  {Karr} J.,  {Su} Y.-N.,  {Liu}
  S.-Y.,  {Froebrich} D.,   {Yeh} C.~C.,  2012, \mn@doi [\apjs]
  {10.1088/0067-0049/200/1/2}, \href
  {http://adsabs.harvard.edu/abs/2012ApJS..200....2L} {200, 2}

\bibitem[\protect\citeauthoryear{{Lee} et~al.,}{{Lee} et~al.}{2013}]{lee_2013}
{Lee} H.-T.,  et~al., 2013, \mn@doi [\apjs] {10.1088/0067-0049/208/2/23}, \href
  {http://adsabs.harvard.edu/abs/2013ApJS..208...23L} {208, 23}

\bibitem[\protect\citeauthoryear{Maret}{Maret}{2015}]{maret_2015}
Maret S.,  2015, thindisk: Thindisk v1.0, \mn@doi{10.5281/zenodo.13823}, \url
  {http://dx.doi.org/10.5281/zenodo.13823}

\bibitem[\protect\citeauthoryear{{Moscadelli} et~al.,}{{Moscadelli}
  et~al.}{2016}]{moscadelli_2016}
{Moscadelli} L.,  et~al., 2016, \mn@doi [\aap] {10.1051/0004-6361/201526238},
  \href {http://adsabs.harvard.edu/abs/2016A%26A...585A..71M} {585, A71}

\bibitem[\protect\citeauthoryear{{Mottram} et~al.,}{{Mottram}
  et~al.}{2011}]{mottram_2011}
{Mottram} J.~C.,  et~al., 2011, \mn@doi [\apjl] {10.1088/2041-8205/730/2/L33},
  \href {http://adsabs.harvard.edu/abs/2011ApJ...730L..33M} {730, L33}

\bibitem[\protect\citeauthoryear{{Olnon}}{{Olnon}}{1975}]{olnon_1975}
{Olnon} F.~M.,  1975, \aap, \href
  {http://adsabs.harvard.edu/abs/1975A%26A....39..217O} {39, 217}

\bibitem[\protect\citeauthoryear{{Ossenkopf} \& {Henning}}{{Ossenkopf} \&
  {Henning}}{1994}]{ossenkopf_1994}
{Ossenkopf} V.,  {Henning} T.,  1994, \aap, \href
  {http://adsabs.harvard.edu/abs/1994A%26A...291..943O} {291, 943}

\bibitem[\protect\citeauthoryear{{Pankonin}, {Churchwell}, {Watson}  \&
  {Bieging}}{{Pankonin} et~al.}{2001}]{pankonin_2001}
{Pankonin} V.,  {Churchwell} E.,  {Watson} C.,   {Bieging} J.~H.,  2001,
  \mn@doi [\apj] {10.1086/322249}, \href
  {http://adsabs.harvard.edu/abs/2001ApJ...558..194P} {558, 194}

\bibitem[\protect\citeauthoryear{{Patel} et~al.,}{{Patel}
  et~al.}{2005}]{patel_2005}
{Patel} N.~A.,  et~al., 2005, \mn@doi [\nat] {10.1038/nature04011}, \href
  {http://adsabs.harvard.edu/abs/2005Natur.437..109P} {437, 109}

\bibitem[\protect\citeauthoryear{{P{\'e}rez} et~al.,}{{P{\'e}rez}
  et~al.}{2012}]{perez_2012}
{P{\'e}rez} L.~M.,  et~al., 2012, \mn@doi [\apjl]
  {10.1088/2041-8205/760/1/L17}, \href
  {http://adsabs.harvard.edu/abs/2012ApJ...760L..17P} {760, L17}

\bibitem[\protect\citeauthoryear{{Qiu}, {Zhang}, {Beuther}  \&
  {Fallscheer}}{{Qiu} et~al.}{2012}]{qiu_2012}
{Qiu} K.,  {Zhang} Q.,  {Beuther} H.,   {Fallscheer} C.,  2012, \mn@doi [\apj]
  {10.1088/0004-637X/756/2/170}, \href
  {http://adsabs.harvard.edu/abs/2012ApJ...756..170Q} {756, 170}

\bibitem[\protect\citeauthoryear{{Rathborne}, {Jackson}, {Chambers},
  {Stojimirovic}, {Simon}, {Shipman}  \& {Frieswijk}}{{Rathborne}
  et~al.}{2010}]{rathborne_2010}
{Rathborne} J.~M.,  {Jackson} J.~M.,  {Chambers} E.~T.,  {Stojimirovic} I.,
  {Simon} R.,  {Shipman} R.,   {Frieswijk} W.,  2010, \mn@doi [\apj]
  {10.1088/0004-637X/715/1/310}, \href
  {http://adsabs.harvard.edu/abs/2010ApJ...715..310R} {715, 310}

\bibitem[\protect\citeauthoryear{{S{\'a}nchez-Monge}
  et~al.,}{{S{\'a}nchez-Monge} et~al.}{2013}]{sanchez_2013}
{S{\'a}nchez-Monge} {\'A}.,  et~al., 2013, \mn@doi [\aap]
  {10.1051/0004-6361/201321134}, \href
  {http://adsabs.harvard.edu/abs/2013A%26A...552L..10S} {552, L10}

\bibitem[\protect\citeauthoryear{{Sanna}, {Reid}, {Carrasco-Gonz{\'a}lez},
  {Menten}, {Brunthaler}, {Moscadelli}  \& {Rygl}}{{Sanna}
  et~al.}{2012}]{sanna_2012}
{Sanna} A.,  {Reid} M.~J.,  {Carrasco-Gonz{\'a}lez} C.,  {Menten} K.~M.,
  {Brunthaler} A.,  {Moscadelli} L.,   {Rygl} K.~L.~J.,  2012, \mn@doi [\apj]
  {10.1088/0004-637X/745/2/191}, \href
  {http://adsabs.harvard.edu/abs/2012ApJ...745..191S} {745, 191}

\bibitem[\protect\citeauthoryear{{Sato} et~al.,}{{Sato}
  et~al.}{2014}]{sato_2014}
{Sato} M.,  et~al., 2014, \mn@doi [\apj] {10.1088/0004-637X/793/2/72}, \href
  {http://adsabs.harvard.edu/abs/2014ApJ...793...72S} {793, 72}

\bibitem[\protect\citeauthoryear{{Shepherd}, {Kurtz}  \& {Testi}}{{Shepherd}
  et~al.}{2004}]{shepherd_2004}
{Shepherd} D.~S.,  {Kurtz} S.~E.,   {Testi} L.,  2004, \mn@doi [\apj]
  {10.1086/380633}, \href {http://adsabs.harvard.edu/abs/2004ApJ...601..952S}
  {601, 952}

\bibitem[\protect\citeauthoryear{{Shimajiri} et~al.,}{{Shimajiri}
  et~al.}{2015}]{shimajiri_2015}
{Shimajiri} Y.,  et~al., 2015, \mn@doi [\apjs] {10.1088/0067-0049/221/2/31},
  \href {http://adsabs.harvard.edu/abs/2015ApJS..221...31S} {221, 31}

\bibitem[\protect\citeauthoryear{{Walsh} et~al.,}{{Walsh}
  et~al.}{2014}]{walsh_2014}
{Walsh} C.,  et~al., 2014, \mn@doi [\apjl] {10.1088/2041-8205/791/1/L6}, \href
  {http://adsabs.harvard.edu/abs/2014ApJ...791L...6W} {791, L6}

\bibitem[\protect\citeauthoryear{{Walsh}, {Nomura}  \& {van Dishoeck}}{{Walsh}
  et~al.}{2015}]{walsh_2015}
{Walsh} C.,  {Nomura} H.,   {van Dishoeck} E.,  2015, \mn@doi [\aap]
  {10.1051/0004-6361/201526751}, \href
  {http://adsabs.harvard.edu/abs/2015A%26A...582A..88W} {582, A88}

\bibitem[\protect\citeauthoryear{{Wang}, {van der Tak}  \&
  {Hogerheijde}}{{Wang} et~al.}{2012}]{wang_2012}
{Wang} K.-S.,  {van der Tak} F.~F.~S.,   {Hogerheijde} M.~R.,  2012, \mn@doi
  [\aap] {10.1051/0004-6361/201117044}, \href
  {http://adsabs.harvard.edu/abs/2012A%26A...543A..22W} {543, A22}

\bibitem[\protect\citeauthoryear{{Wheelwright}, {Oudmaijer}, {de Wit}, {Hoare},
  {Lumsden}  \& {Urquhart}}{{Wheelwright} et~al.}{2010}]{wheelwright_2010}
{Wheelwright} H.~E.,  {Oudmaijer} R.~D.,  {de Wit} W.~J.,  {Hoare} M.~G.,
  {Lumsden} S.~L.,   {Urquhart} J.~S.,  2010, \mn@doi [\mnras]
  {10.1111/j.1365-2966.2010.17250.x}, \href
  {http://adsabs.harvard.edu/abs/2010MNRAS.408.1840W} {408, 1840}

\bibitem[\protect\citeauthoryear{{Wood} \& {Churchwell}}{{Wood} \&
  {Churchwell}}{1989}]{Wood1989apjs}
{Wood} D.~O.~S.,  {Churchwell} E.,  1989, \mn@doi [\apjs] {10.1086/191329},
  \href {http://adsabs.harvard.edu/abs/1989ApJS...69..831W} {69, 831}

\bibitem[\protect\citeauthoryear{{Yorke} \& {Sonnhalter}}{{Yorke} \&
  {Sonnhalter}}{2002}]{yorke_2002}
{Yorke} H.~W.,  {Sonnhalter} C.,  2002, \mn@doi [\apj] {10.1086/339264}, \href
  {http://adsabs.harvard.edu/abs/2002ApJ...569..846Y} {569, 846}

\bibitem[\protect\citeauthoryear{{Zapata}, {Palau}, {Galv{\'a}n-Madrid},
  {Rodr{\'{\i}}guez}, {Garay}, {Moran}  \& {Franco-Hern{\'a}ndez}}{{Zapata}
  et~al.}{2015}]{zapata_2015}
{Zapata} L.~A.,  {Palau} A.,  {Galv{\'a}n-Madrid} R.,  {Rodr{\'{\i}}guez}
  L.~F.,  {Garay} G.,  {Moran} J.~M.,   {Franco-Hern{\'a}ndez} R.,  2015,
  \mn@doi [\mnras] {10.1093/mnras/stu2527}, \href
  {http://adsabs.harvard.edu/abs/2015MNRAS.447.1826Z} {447, 1826}

\bibitem[\protect\citeauthoryear{{de Gregorio-Monsalvo} et~al.,}{{de
  Gregorio-Monsalvo} et~al.}{2013}]{de-gregorio-monsalvo_2013}
{de Gregorio-Monsalvo} I.,  et~al., 2013, \mn@doi [\aap]
  {10.1051/0004-6361/201321603}, \href
  {http://adsabs.harvard.edu/abs/2013A%26A...557A.133D} {557, A133}

\makeatother
\end{thebibliography}

% Don't change these lines
\bsp	% typesetting comment
\label{lastpage}
\end{document}